\documentclass{article}
\usepackage[a4paper, portrait, margin=1.1811in]{geometry}
\usepackage[english]{babel}
\usepackage[utf8]{inputenc}
\usepackage[backend=bibtex,style=nature,sorting=none]{biblatex} 
\usepackage{amsfonts}
\usepackage{amsmath}
\usepackage{amsthm}
\usepackage{amssymb}
\usepackage{appendix}
\usepackage{bbm}
\usepackage{bm}
\usepackage{booktabs}
\usepackage[font={small,it}]{caption} 
\usepackage[autostyle,italian=quotes]{csquotes} 
\usepackage{empheq}
\usepackage{emptypage}
\usepackage{float}
\usepackage{graphicx}
\usepackage{indentfirst}
\usepackage{latexsym}
\usepackage{lettrine}
\usepackage{mathtools} 
\usepackage{microtype} 
\usepackage{physics}
\usepackage{relsize}
\usepackage{siunitx} 
\usepackage{subcaption}
\usepackage{tikz-cd} 
\usetikzlibrary{arrows.meta}
\usepackage{chemfig}
\usepackage{tocloft}
\usepackage{notoccite}
\usepackage{geometry}
\usepackage{fancyhdr} 
\usepackage[version=4]{mhchem}

\DeclareSIUnit{\molar}{M}

\fancypagestyle{plain}{
	\fancyhf{}

}

\graphicspath{ {./Figures/} }

\makeatletter
\patchcmd{\@maketitle}{\LARGE \@title}{\fontsize{16}{19.2}\selectfont\@title}{}{}
\makeatother

\usepackage{authblk}

\newsavebox\affbox
\author[1]{\textbf{Valentina Buonfiglio}}
\author[2]{\textbf{Niccolò Zagli}}
\author[1]{\textbf{Irene Pertici}}
\author[1]{\textbf{Vincenzo Lombardi}}
\author[1]{\textbf{Pasquale Bianco}}
\author[3]{\textbf{Duccio Fanelli}}
\affil[1]{PhysioLab, University of Florence, Sesto Fiorentino (FI), Italy}
\affil[2]{Nordita, Stockholm University and KTH Royal Institute of Technology, Stockholm, Sweden} 
\affil[3]{Department of Physics and Astronomy, University of Florence, Sesto Fiorentino (FI), Italy}

\title{\textbf{\huge Resolving the kinetics of an ensemble of muscle myosin motors via a temperature-dependent fitting procedure}
}

\date{}    

\begin{document}
	
	\pagestyle{headings}	
	\newpage
	\setcounter{page}{1}
	\renewcommand{\thepage}{\arabic{page}}
	
	\maketitle
	
	\noindent\rule{15cm}{0.5pt}
	\begin{abstract}
    A data fitting procedure is devised and thoroughly tested to provide self-consistent estimates of the relevant mechanokinetic parameters involved in a plausible scheme underpinning the output of an ensemble of myosin II molecular motors mimicking the muscle contraction. The method builds on a stochastic model accounting for the force exerted by the motor ensemble operated both in the low and high force-generating regimes corresponding to different temperature ranges. The proposed interpretative framework is successfully challenged against simulated data, meant to mimic the experimental output of a unidimensional synthetic nanomachine powered by pure muscle myosin isoforms.    
		\let\thefootnote\relax\footnotetext{
			\small $^{*}$\textbf{Corresponding authors} \textit{
				\textit{E-mail addresses: \color{cyan}valentina.buonfiglio@unifi.it}}\\
		}
	\end{abstract}
	\noindent\rule{15cm}{0.4pt}

\section{Introduction}

\begin{figure}[t!]
\centering
    \includegraphics[scale=0.15]{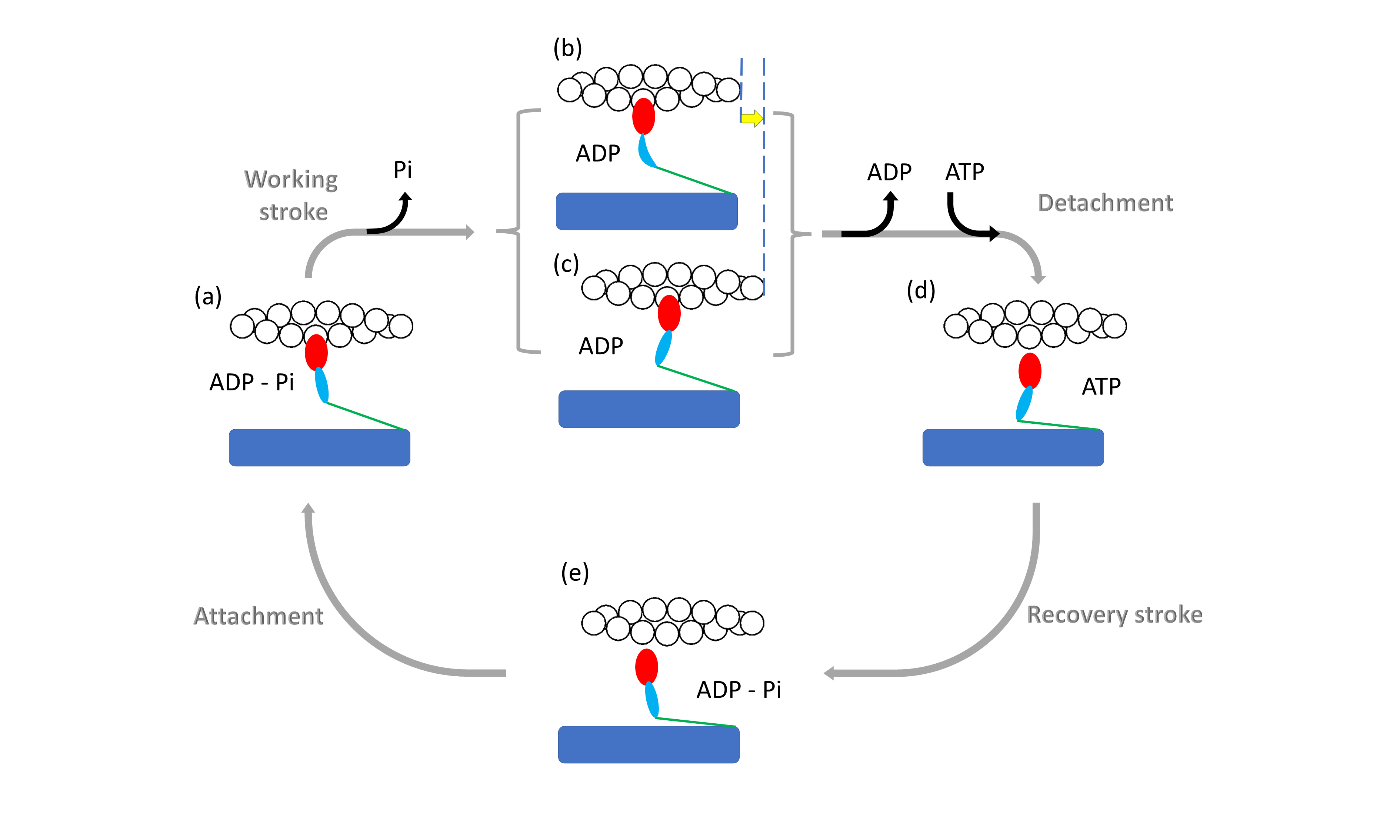}
    \caption{\textbf{Schematic diagram of the chemo-mechanical cycle of the myosin motor during its interaction with the actin filament.} \\
    The HMM fragment of the myosin molecule is a dimer with each monomer made by the subfragment 1 ({\normalfont S1} or head containing the motor domain (red) and the light chain domain (the lever arm, violet)) and the subfragment 2 ({\normalfont S2} or tail, green) extending from the myosin filament backbone (blue). For simplicity, only one {\normalfont S1} and {\normalfont S2} are represented here. 
    The myosin·ADP·Pi complex attaches to actin (white circles) {\normalfont (a)}, forming the cross--bridge, which triggers tilting of the lever arm and Pi release with generation of force and actin filament sliding. 
    If the mechanical load is high, it opposes filament sliding and tilting of the lever arm causes increase in strain in the system, represented here by the distortion of the lever arm {\normalfont (b)}.
    If the load is low {\normalfont (c)} tilting of the lever arm causes actin filaments sliding (yellow arrow), keeping the strain low.
    ADP release from and ATP binding to the motor domain cause myosin detachment from actin. 
    ADP release is slower at high load, ${\normalfont (b)} \to {\normalfont (d)}$, and becomes faster at lower load ${\normalfont (c)} \to {\normalfont (d)}$. 
    Hydrolysis of ATP in the detached head and reversal of the lever arm tilting (recovery stroke, ${\normalfont (d)} \to {\normalfont (e)}$) completes the cross--bridge cycle. 
    The absence of ATP causes the cycle to stop before detachment so that all motors stay attached to actin (rigor). Figure reproduced from \cite{buonfiglio2024force}.} 
    \label{fig:AMcycle}
\end{figure}

Steady force and shortening in the half-sarcomere (the functional unit of the striated muscle cell) are fueled by ATP-driven cyclic interactions of the head (subfragment $1$) of the heavy meromyosin fragment (HMM) of myosin II molecule, extending in array from the thick filament and the actin monomers on the nearby overlapping thin filament. 
The scheme of the above interaction is pictorially depicted in Figure \ref{fig:AMcycle}. 
Following each step, the head undergoes a structural working stroke (a), consisting in a tilting between the motor domain (red) firmly attached to the actin and the light chain binding domain (blue), also called lever arm, connected to the myosin filament backbone through subfragment $2$ (green). 
Under isometric contraction (b), the tilting of the lever arm raises the force exerted by the half-sarcomere, increasing the strain of all the elastic elements. 
When the load is lower than the maximum steady force exerted under isometric conditions, relative filament sliding is instead produced with a reduced strain in the elastic components (c). 
Cyclic asynchronous ATP-driven interactions of myosin motors with the actin filament account for the generation of steady force and shortening. 
Interestingly, the performance of different skeletal muscles and the related energetics rely upon the specificity of the myosin II isoform. Slow muscles, involved in maintenance of posture, are characterised by the dominant presence of the isoform $1$ of Myosin Heavy Chain (MHC$-1$ isoform). 
They exhibit lower shortening speed at any given load, develop lower power and consume ATP at a lower rate, compared to fast muscles. 
These latter, primarily involved in movement, display marked abundance of MHC$-2A$, $-2B$ or $-2X$ isoforms \cite{schiaffino2011fiber}. Strikingly enough, the functional difference between slow and fast isoforms can be traced back to a difference of just $20\%$ in amino-acid composition.

\begin{figure}[!t] 
\centering 
   \includegraphics[scale=0.8]{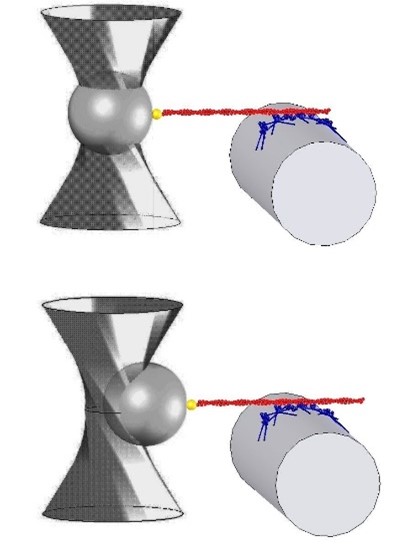}
   \caption{\textbf{Schematic representation of two snapshots during ATP-driven acto-myosin interaction in the nanomachine.} \\
    HMM fragments (blue) are deposited on the nitrocellulose-coated lateral surface of a glass micropipette ( $\sim \SI{3}{\micro \metre}$ diameter, light grey); the $+$end of a single actin filament (red) is attached through gelsolin (yellow) to the bead (dark grey) trapped by the DLOT. Upper panel: the array of motors is brought to interact with the actin filament in solution with $\SI{2}{\milli \molar}$ ATP. Lower panel: once acto-myosin interaction is established, the ensemble of motors starts to develop force up to a steady isometric value ($F_0$). Given the relatively large compliance of the trap, the force generation produces the movement of the bead away from the focus of the DLOT, causing a relative sliding between the actin filament and the motor array.  Figure reproduced from \cite{pertici2018myosin}.}     
\label{fig:dlot}
\end{figure}

\begin{figure}[!t] 
\centering 
   \includegraphics[scale=0.6]{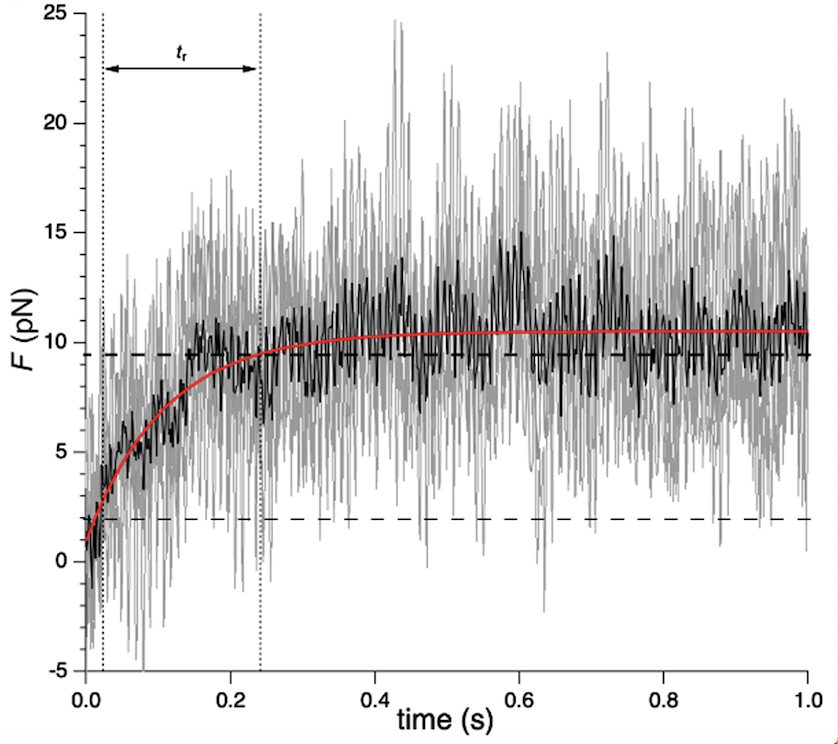}
   \caption{\textbf{Time course of force development for a slow myosin (soleus) array.} \\
   The curves refer to $6$ records from independent experiments (grey traces). The black line is obtained by averaging over distinct realization of the stochastic dynamics. The red line is a guide for the eye drawn from a single exponential fitting of the averaged data ($t_r$ quantifies the estimated raising time).}     
\label{fig:output}
\end{figure}

\noindent
Recording actin-myosin interaction \textit{in vitro} via single molecule mechanics does not give access to the emergent properties of the ensemble of motors acting in the half—sarcomere \cite{finer1994single, ishijima1996multiple}. 
Recently, a unidimensional synthetic nanomachine was developed \cite{pertici2018myosin} which allows for significant progress to be made in the exploration of motor ensemble kinetics. 
Thanks to this technology, one can address the generation of steady force and the shortening by an array of pure myosin isoforms interacting with the actin filament in a controlled framework, in the absence of the contribution of the other sarcomeric proteins and higher hierarchical levels of organisation of the muscle. 
More specifically, in the nanomachine a few HMM fragments extending from the functionalised surface of a micropipette carried on a three-way nanopositioner (acting as a length transducer) are brought to interact with an actin filament attached with the correct polarity to a bead trapped by a Dual Laser Optical Tweezers system, DLOT (acting as a force transducer)(Figure \ref{fig:dlot}).  
Following the nanomachine implementation described in \cite{buonfiglio2024force}, the system can be operated either in position clamp condition, when the feedback signal is the position of the motor array support, or in length clamp condition, when the feedback signal is the difference between the position of the bead and that of the motor support. 
The length clamp suppresses the sliding between the actin filament and the array of motors caused by force generating interactions in the presence of the intrinsically large trap compliance. 
In length clamp any displacement of the bead is counteracted by a displacement of the motor support carried by the nanopositioner. 
As detailed in \cite{buonfiglio2024force}, under this condition motors act as independent force generators and the rate of development of the steady isometric force, as well as the  superimposed fluctuations, are a direct expression of the attachment/force-generation/detachment kinetics of motors.
\noindent
In Figure \ref{fig:output} a typical record of the force developed during an interaction of the actin filament with the motor array (rise and saturation of the collective force) is displayed. 
Additional details on the experimental apparatus are provided in \cite{buonfiglio2024force}.
\noindent
The data collected from myosin isoforms purified from slow and fast muscle in \cite{buonfiglio2024force} were used to feed a stochastic model. 
For each isoform the force of individual motors, denoted with $f_0$, the fraction of actin-attached motors at the force plateau, $r$, and the rate of transition through the attachment-detachment cycle, $\phi$, were successfully estimated via a newly designed fitting protocol which exploits the information stored in the fluctuating component of the force. 
To handle the technicalities involved in the  proposed fitting scheme, it was assumed that the contribution to the force (including fluctuations) of the motors in the state $A_1$ (i.e. attached motors in the low-generating force regime) is always negligible, a working ansatz that, according to the literature \cite{piazzesi2003temperature,decostre2005effect}, is correct as far as the temperature of the investigated sample is sufficiently high. 
Multiple combinations of the involved kinetic parameters however exist that yield the same fitted profile, with almost identical estimates for the relevant quantities, including the aforementioned parameters $f_0$, $r$ and $\phi$. 
This conclusion was reached by challenging the proposed method against synthetically generated data, that bear a degree of complexity similar to what reflected in their experimental homologue. 
Starting from these premises, we here set to explore a non trivial extension of the fitting strategy discussed in \cite{buonfiglio2024force}, which allows for multiple temperatures dataset to be simultaneously analysed, with explicit inclusion of the contribution to the force as stemming from the motors in the low-generating force regime and thus extending to ideally embrace the low temperature regime. 
As we shall prove by testing the proposed methodology against suitably tailored simulated data, we can successfully estimate all the kinetics, \textit{a priori} unknown, parameters of the mechanochemical cycle with unprecedented accuracy and without any assumption from muscle or cell studies. 
In the following, we set to illustrate the basic of the method and report on the implemented tests.

\section{Minimal stochastic model for acto-myosin interactions}
\noindent
The molecular basis for the force generation in the skeletal muscle is investigated on a simplified one-dimensional system constituted by a single actin filament, interacting with an ensemble of $N$ myosin motors.
In the presence of physiological ATP concentration, the system develops the steady force typical of muscle isometric contraction.  
In order to provide a description of the force-generating interactions between the motor ensemble and the actin filament, we will consider a simple reaction scheme that allows for a minimal non equilibrium description of the actin-myosin interactions. \\
We assume that each motor of the ensemble can be found in three different configurations: a detached state $D$, resulting in a vanishing force, and two different force-generating attached states $A_1$ and $A_2$. 
In particular, $A_1$ ($A_2$) corresponds to a low (high) force configuration (see later discussion). 
We model each cyclical actin-myosin interaction by the following kinetic scheme:
\begin{equation}
\label{eq: chemical equations}
	\ce{ D <=>[\ce{\ k_1}\ ][\ce{\ k_{-1}}\ ] A_1 <=>[\ce{\ k_2 \ }][\ce{\ k_{-2}}\ ] \ce{A_2} <=>[\ce{\ k_3 \ }][\ce{\ k_{-3}}\ ] D }
\end{equation}
where the $k_i >0$, with $i \in \{\pm 1, \pm 2, \pm3 \}$, represent the rates associated to each step in the reaction cycle.
The interpretation of the above chemical equations is straightforward. Considering as an example the first step, writing $ \ce{ D ->[\ce{\ k_1}\ ]A_1 }$ implies that there is a probability (per unit of time) equals to $k_1$ for a motor in the detached state $D$ to transition to the attached state $A_1$. 
In the following we will consider operating regimes of the ensemble of the motors corresponding to negligible values of the rate determining the reverse transition from the detached state $D$ to attached state $A_2$ and then we will then set $k_{-3}=0$. 
We remark that this unequivocally sets the system described by equations \eqref{eq: chemical equations} to be in non-equilibrium conditions where the ATP concentration is kept constant at all times while the ATP hydrolysis products are continually delivered. 
We refer the reader to Appendix \ref{App:TD} for a detailed thermodynamic description of the kinetic scheme \eqref{eq: chemical equations}. \\
The state of the ensemble of the motors can be fully described at the mesoscopic level by considering the number $n_D$ of detached motors and by the number $n_1$ ($n_2$) of attached motors in configurations $A_1$ ($A_2$). Considering that the total number $N$ of motors does not vary with time, it is possible to write the conservation law $N = n_D + n_1+ n_2$.
As a consequence, the mesoscopic state of the system can be described solely by the two dimensional vector $\mathbf{n}\left(t\right)= \left(n_1\left(t \right), n_2\left(t\right) \right)$. 
The stochastic evolution of the population $\mathbf{n}(t)$ is determined by the chemical equations \eqref{eq: chemical equations} and can be simulated using numerical routines for stochastic jump processes such as the popular Gillespie algorithm \cite{gillespie1976general,gillespie1977exact}. 
Alternatively, the statistical properties of the system can be described in terms of a master equation. 
In this direction, we denote with $P(\mathbf{n},t)$ the probability of observing the system in configuration $\mathbf{n}$ at time $t$ and with $T(\mathbf{n}|\mathbf{n}')$ the transition rate (probability per unit of time) from the configuration $\boldsymbol{n'}$ to a new configuration $\boldsymbol{n}$. The master equation determines the evolution in time of $P(\mathbf{n},t)$ and can be written as
\begin{equation}
\label{eq: ME}
    \frac{\partial P(\boldsymbol{n},t)}{\partial t}= \sum_{\boldsymbol{n'}\ne \boldsymbol{n}} \Bigl[ T(\boldsymbol{n} \vert \boldsymbol{n'}) P(\boldsymbol{n'},t)- T(\boldsymbol{n'} \vert \boldsymbol{n})P(\boldsymbol{n},t) \Bigr] \    
\end{equation}
We remark that the master equation represents a balance of opposite contributions: the transitions {\it towards} the reference state (the associated terms bearing a plus sign), and the transitions {\it from} the reference state (terms with a minus). 
According to the reaction cycle \eqref{eq: chemical equations} one can identify the following transition rates between states of the population.
\begin{itemize}
    \item Attachment process\footnote{In order to identify the arrival/departure state $\boldsymbol{n'}$ we only highlight the individual component that changes according to the considered reaction}: $T(n_1+1 \vert \boldsymbol{n})= k_1 \ \frac{n_D}{N}= k_1 \left( 1- \frac{1}{N} \left(n_1+n_2\right)\right)$
    \item Detachment from configuration $A_1$ process:  $T(n_1-1 \vert \boldsymbol{n})= k_{-1} \ \frac{n_1}{N}$
    \item Detachment from configuration $A_2$ process:  $ T(n_2-1 \vert \boldsymbol{n})= k_3 \ \frac{n_2}{N}$
    \item Conversion $A_1 \to A_2$ process:  $T(n_1-1, n_2+1 \vert \boldsymbol{n})= k_{2} \ \frac{n_1}{N}$
    \item Conversion $A_2 \to A_1$ process: $T(n_1+1, n_2-1 \vert \boldsymbol{n})= k_{-2} \ \frac{n_2}{N}$
\end{itemize}
We remark that the long time statistics of the system, attained when transient effects have passed, are fully encoded in the stationary solution $P_{st}(\mathbf{n})$ of the master equation that is obtained by setting to zero the r.h.s. of equation \eqref{eq: ME}.
\\
The above equations provide an exact mathematical description of the statistical properties of the evolution of the population of motors. 
However, the microscopic dynamics of the population of motors is not directly experimentally accessible. We will then aim to characterise the total force exerted by the ensemble of motors in isometric conditions. 
According to the design of the nanomachine \cite{pertici2018myosin}, it is necessary to assume motors have a random orientation with respect to the actin filament.
As a consequence, we assume that the force exerted by a motor depends on the binding angle $\theta$, as measured from the correct orientation with respect to the actin filament, corresponding to the \textit{in situ} orientation in the half-sarcomere. 
The largest value of the force $f_0$ is exerted when the motor orientation is correct ($\theta=0$).
Depending on the specific orientation of the motor, the force progressively decreases up to a minimum value $f_0/10$ ~\cite{ishijima1996multiple}. 
In accordance with \cite{pertici2020myosin,buonfiglio2024force}, we postulate that the exerted force $f_1$ by each motor in configuration $A_1$ (low-force) is a random variable, uniformly sampled from the bounded interval $\mathcal{I}_1=\Bigl[-f_0, f_0\Bigr]$ every time a motor gets into such configuration, that is $f_1 \sim U(-f_0,f_0)$. Similarly, the force $f_2$ that each motor in configuration $A_2$ (high-force) is uniformly distributed $f_2 \sim U(\frac{f_0}{10}, f_0 )$. 
The total force generated by the ensemble in state $\mathbf{n}$ is then
\begin{equation}
\label{eq: definition of the force}
    F_{n_1, n_2}= \sum_{i=0}^{n_1} f_1^{(i)}+ \sum_{i=0}^{n_2} f_2^{(i)}
\end{equation}
where the index $i=1,\dots,n_1$ $(n_2)$ identifies each motor in configuration $A_1$ ($A_2$). 
We remark that $F_{n_1,n_2}$ is a random variable whose statistical properties originate from the randomness due to two contributions: on the one side, the stochastic evolution of the population dynamics given by the master equation, and, on the other side, the randomness associated to the single motor forces. 
In particular, we are interested in evaluating the statistical properties of the variable $F_{n_1,n_2}$ both during the force development and at the plateau, that is once the system has reached statistical steady state.

\subsection{Neglecting fluctuations: force development and isometric plateau}
\noindent \label{Sec:MF}
As a first order approximation, we here approach the analysis of the properties of $F_{n_1,n_2}$ by neglecting fluctuations, that is by considering only the first moment $\mathbb{E}\big[ F_{n_1,n_2} \big] \coloneqq F(t)$ of the total force of the ensemble. Considering  \eqref{eq: definition of the force}, it is easy to see that
\begin{equation}
\label{eq: First moment force}
    F(t) = \langle n_1 \rangle(t) \bar{f}_1 + \langle n_2 \rangle(t) \bar{f}_2 = \frac{11}{20} f_0 \langle n_2 \rangle(t)  \coloneqq  \frac{11}{20}  N f_0 z(t)
\end{equation}
where $\bar{f}_i$ corresponds to the average value of the random force $f_i$ and $\langle \cdot \rangle$ defines the expectation value over the population distribution $P(\mathbf{n},t)$. 
In writing the previous equation we have assumed that $\bar{f}_1 =0$, $\bar{f}_2= \frac{11}{20}f_0$ and we have defined the mean field concentrations  $y(t)= \langle n_1 \rangle/N$ and $z(t)=\langle n_2 \rangle/N$. 
Equation \eqref{eq: First moment force} shows that, if fluctuations are neglected, the motors in configuration $A_1$ do not contribute directly to the total average force as $\bar{f}_1 = 0$. 
As shown in \cite{piazzesi2003temperature}, if high temperature regimes are considered, they do not contribute at all to the average force. 
However, in general conditions, their presence will affect the concentration $z(t)$ of high force producing motors as prescribed by the following equations \eqref{eq: ensemble isometric force} and \eqref{eq: mean field at any time}. \\
The equations that determine the evolution of the average concentrations is easily obtained with standard techniques from the master equation, e.g. by multiplying \eqref{eq: ME} by $n_i$ and sum over all possible values of $n_i$. The resulting system of equations for the concentration vector $\mathbf{x} = (y,z)^T$ is 
\begin{equation}
\label{eq: deterministic equations concentrations}
    \frac{\mathrm{d} \mathbf{x}}{\mathrm{d}t}
= \mathbf{J} \mathbf{x} + \mathbf{b}
\end{equation}
where $\mathbf{b} = (k_1,0)^T$ and  $\mathbf{J}$ is the constant Jacobian matrix of the linear system
\begin{equation}
	\label{Jmatrix}
	\mathbf{J}=
	\begin{pmatrix}
		-(k_1+k_{-1}+k_2) & -k_1+k_{-2} \\
		k_2 & -(k_{-2}+k_3)
	\end{pmatrix}
\end{equation}
Solutions of the above system of equations are related to the spectral properties of $\mathbf{J}$. 
In particular, we denote with $\lambda_{1,2}$ the eigenvalues of the matrix and with $\bm{v}_{1,2}$ the corresponding eigenvectors. 
We remark that $\forall k_i >0$ the eigenvalues of the matrix have negative real part, $\lambda_i <0$, meaning that the system is stable. 
In the considered dynamic regimes all $\lambda_i$ are real.  
In particular, the steady state regime of \eqref{eq: deterministic equations concentrations} is the stable fixed point $\mathbf{x}^* = (y^*,z^*)^T =  \mathbf{J}^{-1}\mathbf{b}$. The (mean) force at plateau, $F_0$, is then 
\begin{equation}
\label{eq: ensemble isometric force}
    F_0 = \frac{11}{20}N f_0 z^* = \frac{11}{20}N f_0 \frac{k_1}{k_1+G}\frac{k_2}{k_2+k_{-2}+k_3}
\end{equation}
with $\displaystyle{G= \frac{k_{-1}(k_{-2}+k_3)+k_2k_3}{k_2+k_{-2}+k_3}}$, where we have explicitly written the rather lengthy expression of $z^*$ in terms of the kinetic rates.  
Other quantities of interest are the duty ratio $r$ of the ensemble of motors, i.e average fraction of actin-attached motors, and the rate of transition through the attachment-detachment cycle $\phi$: 
\begin{align}
\begin{split}
    r &= y^* + z^*, \\
    \phi &= k_3 z^*.
\end{split}
\end{align}
Equation \eqref{eq: deterministic equations concentrations} also characterises the mean force development profile. As a matter of fact, due to the linearity of the system, the solutions can be analytically obtained as:
\begin{equation}
\label{eq: mean field at any time}
    \mathbf{x}(t) = \mathbf{x}^* + c_1 e^{-|\lambda_1| t} \mathbf{v}_1 + c_2 e^{-|\lambda_2| t} \mathbf{v}_2
\end{equation}
where the $c_i$s are imposed by the initial conditions. The experimental protocol we want to describe is characterised by having all the motors being in an initial detached configuration, equivalent to setting $\mathbf{x}(t=0) =(0,0)^T$. The initial conditions are then found by solving the system of equations 
\begin{equation}
\label{eq: initial conditions mean field}
    \mathbf{x}^* + c_1 \mathbf{v}_1 + c_2 \mathbf{v}_2 = 0
\end{equation}
Equation \eqref{eq: First moment force} together with \eqref{eq: mean field at any time} and \eqref{eq: initial conditions mean field} describe the average force exerted by the ensemble of motors at any time $t$, including the initial force development process. 
We remark that in \cite{buonfiglio2024force} it was assumed a high temperature regime where the effect of the motors in configuration $A_1$ was negligible, giving a duty ratio $r \approx z^*$. 
Here we do not make such assumption and let motors in both force-producing configurations to be at play. In particular, this results in having two comparable timescales $\tau_i = \frac{1}{|\lambda_i|}$ shaping the dynamics of the force development process. 
Furthermore, we observe that, even if both force producing configurations contribute to the duty ratio of the system, only motors in the high-force producing configuration $A_2$ will contribute directly to the mean force at plateau $F_0$. In order to extract information on both force producing configurations it is then necessary to consider into the analysis the fluctuations around $F_0$.

\subsection{Going beyond the first moment: the full distribution of the ensemble force}
\label{sec: Distribution of force}
\noindent
We here evaluate the distribution of the total force $F_{n_1,n_2}$ at the plateau, that is when the statistical properties of the population dynamics are described by the stationary solution $P_{st}$ of the master equation \eqref{eq: ME}. We hence want to evaluate the probability $P(F_{n_1,n_2}=F)$ that the random variable $F_{n_1,n_2}$ is equal to $F$. By the law of total probability we can write that 
\begin{equation}
\label{eq: first def distribution of force}
    P(F_{n_1,n_2}=F)= \sum_{q_1, q_2=0}^N P(F_{q_1,q_2}= F) P_{st}(\mathbf{n}=(q_1,q_2))
\end{equation}
The previous equation should be interpreted as follows. The distribution $P(F_{n_1,n_2}=F)$ is obtained as the sum of the force distributions $P(F_{q_1,q_2}= F)$ over all the possible fixed configurations of population dynamics $\mathbf{n}=(q_1,q_2)$, each weighted with their occurrence probability $P_{st}(\mathbf{n}=(q_1,q_2))$ given by the stationary solution of the master equation. We recall that, fixed a number $q$, the variable $\sum_{i=0}^q f_{1,2}^{(i)}$ is distributed according to a generalisation of the Irwin-Hall distribution \cite{hall1927distribution}. Hence, the probability distribution $P(F_{q_1,q_2}= F)$ of the random variable
\begin{equation}
    F_{q_1,q_2} = \sum_{i=0}^{q_1} f_1^{(i)}+ \sum_{i=0}^{q_2} f_2^{(i)}
\end{equation}
is the distribution of the the sum of two random variables each one of them distributed according to a Irwin-Hall distribution. 
As we want to leverage the knowledge of $P(F_{n_1,n_2}=F)$ to extract information on the parameters of the system through a fitting procedure, see section \eqref{Sec:fit2motors} for more details, we wish to obtain an expression of $P(F_{n_1,n_2}=F)$ that is amenable to a numerically swift and easy computation and that guarantees a simple interpretation too. 
As a matter of fact, computing the distribution of a sum of Irwin-Hall distributions for all configurations $\mathbf{n} = (n_1,n_2)$ results in a quite slow and inefficient way, especially when introduced in a fit procedure. \\
It is possible to find a much less computationally demanding approximation of the total force distribution when we assume that the most likely configurations of the population correspond to states featuring a non-small number of attached motors. Mathematically, we assume that  $P_{st}(\mathbf{n}=(q_1,q_2)) > 0 $ only if $ q_1,q_2 \gg 1$. Essentially, this allows us to approximate any of the Irwin-Hall distributions in equation \eqref{eq: first def distribution of force} as a Gaussian distribution with suitable parameters (according to the Central Limit Theorem). \\
Denoting with $\Phi(n; 0,1)$ the Irwin Hall distribution generated by the sum of $n$ terms of i.i.d. variables uniformly distributed on the unit interval $(0,1)$, we recall that for sufficiently large $n$:
	\begin{equation}
		\Phi(n; 0,1) \xrightarrow{n \gg 0} \mathcal{G}\Big(\frac{n}{2}, \frac{n}{12}\Big) \quad .
	\end{equation}
where $\mathcal{G}(\mu,\sigma^2)$ is the Gaussian distribution with mean $\mu$ and variance $\sigma^2$.
Since any random variable $f_i$ uniformly distributed over an interval $[a,b]$, that is $f_i \sim U(a,b)$, can be written as $f_i=a- (b-a)u_i$ where $u_i \sim U(0,1)$ it is simple to see that when $n \gg 1$  
\begin{equation}
    \sum_{i=1}^n f_i  \sim \mathcal{G}(\frac{n}{2}(a+b),\frac{n}{12}(b-a))
\end{equation}
Considering that the random force $f_1^{(i)}$ ($f_2^{(i)}$) expressed by motors in configuration $A_1$ ($A_2$) is distributed according to a uniform distribution $U(-f_0,f_0)$ ($U(\frac{f_0}{10},f_0)$ ), we conclude that the distribution of the force $F_{q_1,q_2}$ given by a fixed configuration $\mathbf{n}= (q_1,q_2)$ is the one resulting from the sum of two independent normally distributed random variables and it is thus a Gaussian distribution itself given by
	\begin{equation}
		F_{q_1,q_2} \sim \mathcal{G}(\mu_F, \sigma_F^2) \quad \text{with}\\ 
		\begin{cases}
			\mu_F = \displaystyle{\frac{11}{20} q_2 f_0}\\
			\\
			\sigma_F^2 =  \displaystyle{ \big(\frac{1}{3} q_1 + \frac{27}{400}q_2 \bigr) f_0^2} \\
		\end{cases}
	\end{equation}
We remark that the parameters $\mu_F$ and $\sigma_F^2$ depend on the force of single motor $f_0$ and by the population state $\mathbf{n}=(q_1,q_2)$. 
It is interesting to note that the mean $\mu_F$ depends only on the number of motors $q_2$ in the high-force configuration $A_2$ whereas the variance depends both on $q_1$ and $q_2$. This confirms the statement that, if fluctuations are ignored, no information is gained regarding the configuration $A_1$. 
In conclusion, we propose to approximate the distribution of the total force of the ensemble of motors at the plateau by  a sum of suitably weighted Gaussian distributions as 
\begin{equation}
		\label{eq: PF_n1n2}
		P(F)= \sum_{q_1, q_2=0}^N \mathcal{G}\bigl(\mu_F(q_1,f_0), \sigma_F^2(q_1,q_2,f_0) \bigr) P_{st} \bigl(\mathbf{n}=(q_1,q_2) \bigr)
	\end{equation}
We observe that we have used a slightly different notation for the distribution of the force which is now coherent with the notation for a  distribution of a continuous random variable. This is due to the fact that the assumption on the stationary distribution of the master equation implies, \textit{de facto}, a continuous limit for the total force of the ensemble. In Appendix \ref{App:VK} we propose a further simplification of the probability distribution, based on a continuum limit of $P_{st}$  valid in the $N \to \infty$ regime. 
In the following sections we will however use equation \eqref{eq: PF_n1n2} to fully resolve the intrinsic discreteness of the problem.

\subsection{Parameter estimation of the mechanokinetic properties}   \label{Sec:fit2motors}
\begin{figure}[t!]
\begin{subfigure}{0.5\textwidth}
     \includegraphics[scale=0.45]{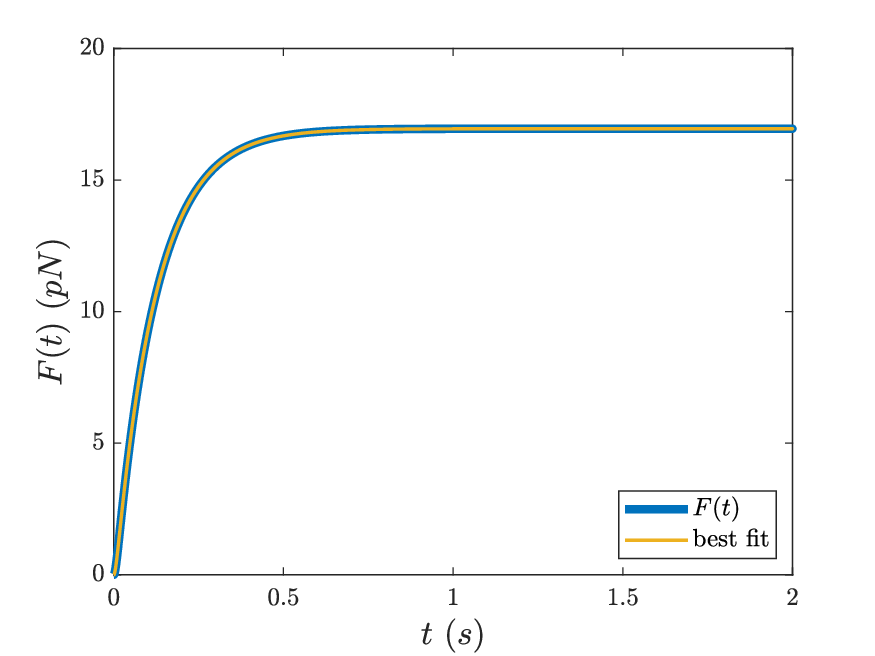}
\subcaption{}
\end{subfigure}
\begin{subfigure}{0.5\textwidth}
    \includegraphics[scale=0.45]{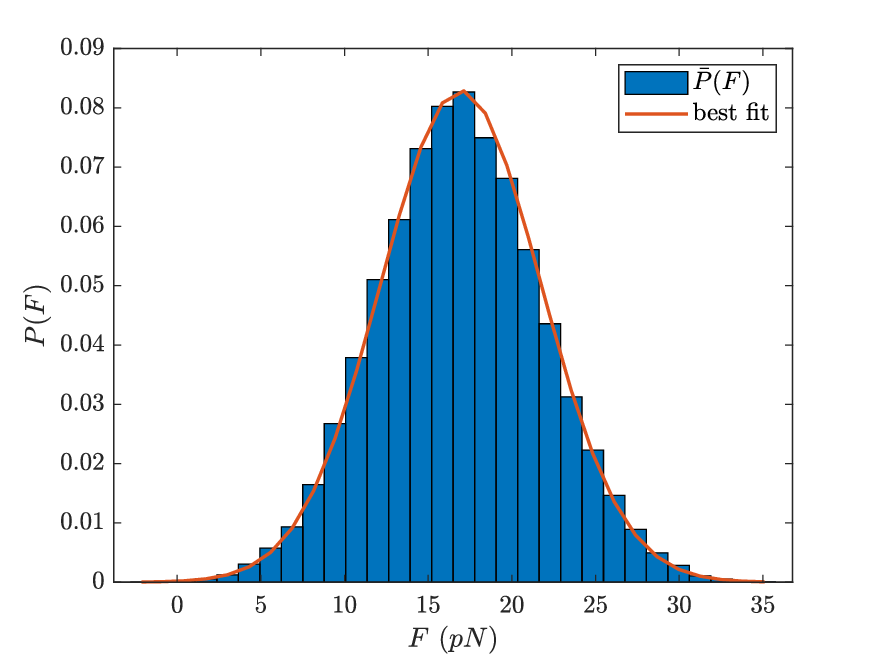}
\subcaption{}
\end{subfigure}
    \caption{\textbf{Result of the fitting procedure on the force development and on the probability distribution of the force obtained from the stochastic simulations of the dynamics.}\\
    \textnormal{\textbf{a.}} The reference force trajectory to be interpolated is here assumed in its mean filed version, by neglecting the finite size corrections.
    The force is measured in $\SI{}{\pico \newton}$ and it is exerted by a collection of $N = 20$ molecular motors with a suitable choice of the kinetic parameters and the force of a single motor $f_0$.
    \textnormal{\textbf{b.}} Comparison of the empirical distribution $\bar{P}(F)$ (histogram, in blue) and the analytical one $P(F)$ (line, in red). }
 \label{fig:fitData}
\end{figure}

\begin{figure}[t!]
    \centering
    \includegraphics[scale=0.6]{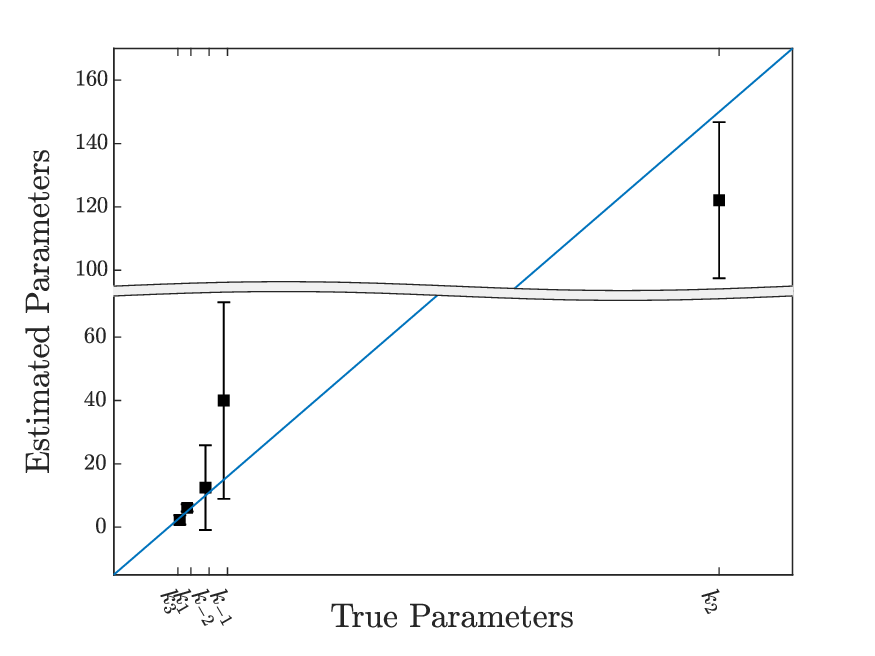}
    \caption{\textbf{Estimated kinetic parameters from the synthetic data.}\\
    The solid blue line represents the true parameters used the generate the synthetic data. The symbols (mean $\pm$ SD) are the estimates obtained with the optimization procedure.}
\label{fig:parBestVSparTrue}
\end{figure}

\noindent
We now present the fitting procedure adopted to perform a parameter estimation of the relevant mechanokinetic parameters using synthetically generated data. The overall goal is to not only estimate the single motor force $f_0$ and the ensemble duty ratio $r$, but also the microscopic kinetic rates $k_i$, $i=\{\pm 1, \pm 2,3\}$ shaping the stochastic dynamics given by equations \eqref{eq: chemical equations}.
The estimation procedure makes use of the information coming from the average force development process described by equation \eqref{eq: First moment force} and the distribution, at plateau, of the force generated by the ensemble as described by equation \eqref{eq: PF_n1n2}.
Firstly, the synthetic data is generated via a Gillespie algorithm whose statistical properties are in accordance with the master equation stationary solution of the motors populations, and with the theoretical expression of the stationary probability distribution \eqref{eq: PF_n1n2}. 
In order to estimate the stationary statistical properties of the force, we remove a fixed period making sure to be safely after the transient force development process and then compute the empirical distribution $\bar P(F)$. 
The reference force trajectory to be interpolated is here assumed in its mean field version, by neglecting the finite size corrections, see eq. \eqref{eq: First moment force}. 
This formally amounts to averaging individual stochastic trajectories over a sufficiently large, virtually infinite, ensemble of realisations. This working ansatz can be relaxed yielding to qualitatively similar conclusions.
The fit is based on a simulated annealing algorithm to optimise the loss function
\begin{equation}
    \mathcal{L} = \gamma_1 |\bar{F}(t) - F(t)|^2 + \gamma_2 |\bar{P}(F) - P(F)|^2
\end{equation}
where $|\cdot|$ represents the usual $L_2$ (quadratic) norm and $\gamma_{1,2}$ are suitable scaling hyper-parameters. 
A graphic result of the fitting procedure of the force development and the force distribution at the isometric plateau is shown in Figure \ref{fig:fitData}, (a) and (b) respectively, while the values of the parameters obtained are listed in Table \ref{tab:parSim}.  
\begin{table}[h!]
	\begin{center}
		$\begin{array}{ccccccc}
			\toprule
			\text{} & F_0 \ (\SI{}{\pico \newton}) & f_0\ (\SI{}{\pico \newton}) & r & \phi (\SI{}{\second^{-1}})
			\\
			\midrule
			\text{True}& 16.9 & 3.0 & 0.56 & 1.5 &
			\\
			\text{parameters }&  &    &     & &\\
			& & & & & \\
			\text{Estimated }& 16.9 & 3.1 \pm 0.3 & 0.56 \pm 0.09 & 1.1 \pm 0.7  &
			\\
			\text{parameters }&   &   &     & &\\
			\bottomrule
		\end{array} $
  \caption{\textbf{Estimated parameters via the inverse scheme fed with simulated data.}\\
			The parameters are: the force of a single motor $f_0$, the duty ratio of the ensemble $r$ and the rate of transition through the attachment–detachment
			cycle $\phi$. Mean and standard deviations are computed from different independent realisation of the optimisation procedure. 
		}
		\label{tab:parSim}
		
	\end{center}
\end{table}
From the inspection of Table \ref{tab:parSim} it can be appreciated that the estimates of $f_0$, $r$ and $\phi$ are in good agreement with the values of the corresponding parameters set when we generated the synthetic data. We remark that the optimisation procedure is quite stable with respect to these parameters, as suggested by the small standard deviations obtained from different independent realisation of the fitting procedure.
On the other hand, the values of the kinetic parameters $k_i$ are not resolved in a satisfactory way by the optimisation, see Figure \ref{fig:parBestVSparTrue}. 
This is due to the large degeneration of possible kinetics solutions corresponding to the same stationary distribution $P(F)$, similarly to what has been previously observed in \cite{buonfiglio2024force} for another working regime.

\section{Temperature dependent parameter estimation strategy} \label{Sec:temp}
\begin{figure}[th!]
\begin{subfigure}{0.5\textwidth}
     \includegraphics[scale=0.45]{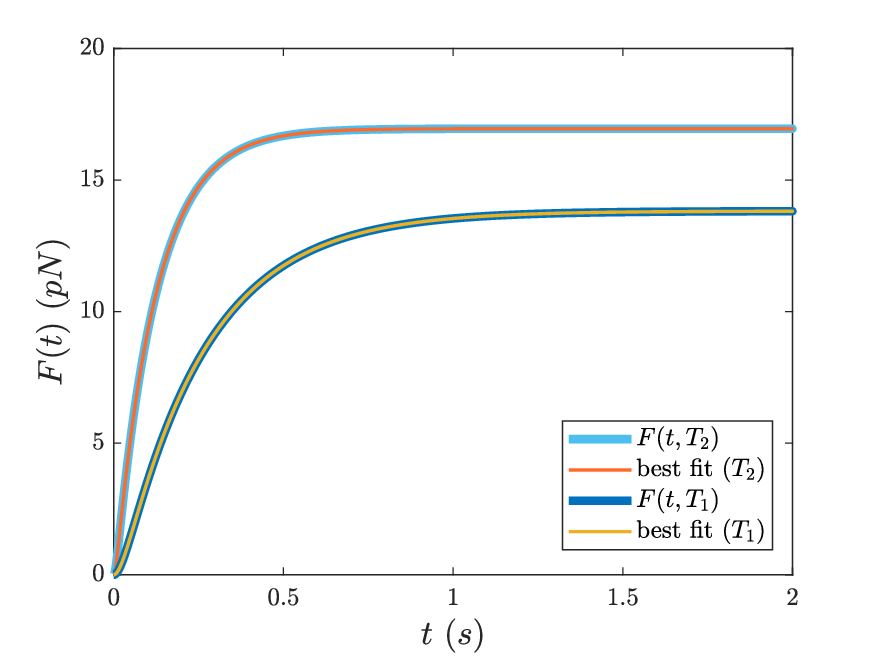}
\subcaption{}
\end{subfigure}
\begin{subfigure}{0.5\textwidth}
    \includegraphics[scale=0.45]{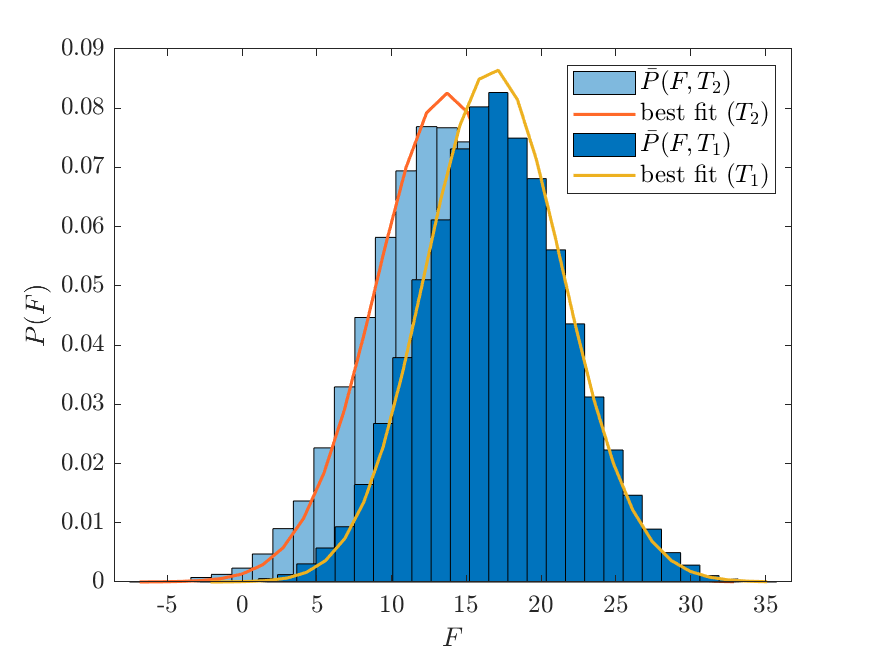}
\subcaption{}
\end{subfigure}
    \caption{\textbf{Result of the fitting procedure on the force development and on the probability distribution of the force obtained from the stochastic simulations of the dynamics.}\\
    The relevant parameters of the simulations are: $N= 20$, $f_0= 3, T_1= \SI{25}{\degreeCelsius}$ and $T_2= \SI{15}{\degreeCelsius}$, while the values of the $Q_{10}$ factors are listed in the text, and the kinetic constants are: 
    $k_1= 5$, $k_{-1}= 15$, $k_2= 150$, $k_{-2}= 10$ and $k_3= 3$.
    \textnormal{\textbf{a.}} The fitted profile of the force refers to its mean field description. 
    \textnormal{\textbf{b.}} The histogram of the force of the ensemble at the isometric plateau is fitted against the analytical profile, via a self–consistent optimisation procedure which aims at estimating the kinetic parameters and the force of a single motor $f_0$.} 
 \label{fig:fitDataT1T2}
\end{figure}

\begin{figure}[!t]
    \centering
    \includegraphics[scale=0.6]{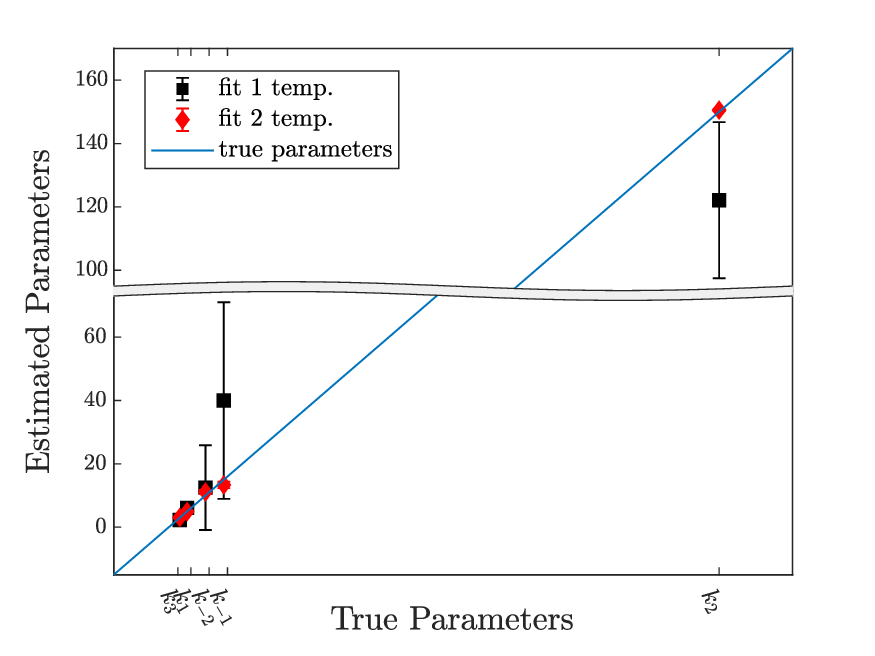}
    \caption{\textbf{Estimated parameters from the synthetic data.} \\
    The solid blue line represents the true parameters of the simulated dynamics, the red symbols (mean $\pm$ SD) are obtained with the optimisation procedure that employs two force data sets at different temperature parameters. The Black symbols (mean $\pm$ SD) report the best fit values as obtained with just one temperature data set (see Figure \ref{fig:parBestVSparTrue}).
    The results shown are relative to the data set at the temperature $T_1= \SI{25}{\degreeCelsius}$.}
	\label{fig:parBestVSparTrue_2temp-lin}
\end{figure}

\begin{figure}[!t]
	\centering
	\includegraphics[scale=0.6]{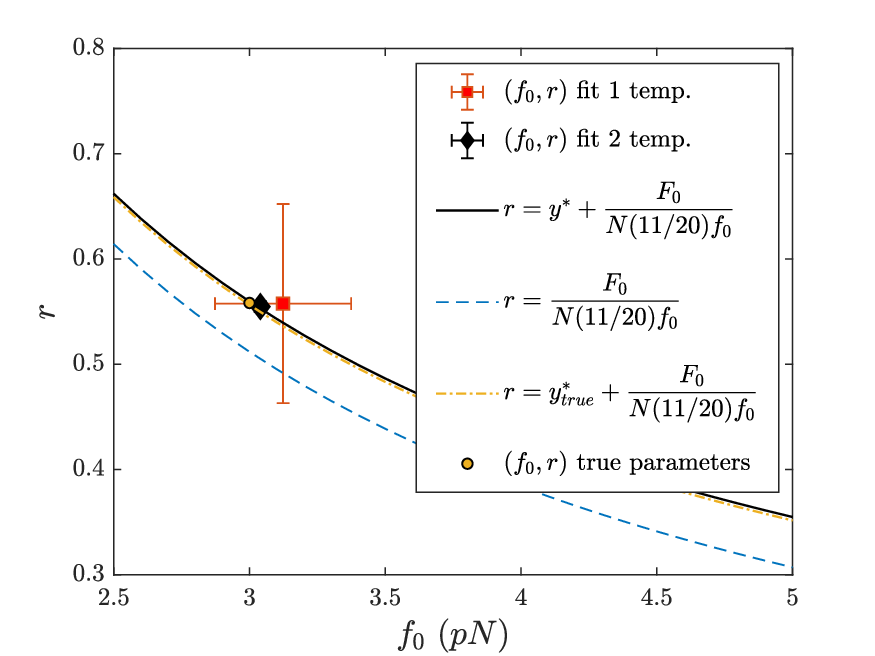} 
	\caption{\textbf{Estimated $\bm{(f_0,r)}$ from the synthetic data.}\\
	The red (black) symbol (mean $\pm$ SD) refers to the fitting operated with the two (one) data sets.
        The solid red line represents the expression of the duty ratio of the ensemble as it follows from the mean field analysis, with the mean field solution for $y^*$ expressed as a function of the estimated kinetic parameters of the system (from the generalised two temperature setting). 
        The dashed blue line represents the duty ratio estimated by using expression following the adiabatic approximation from \cite{buonfiglio2024force}, where the contribution of the motors in configuration $A_1$ is neglected.
        The yellow circle and line are the corresponding data from the original simulation.
	}
	\label{fig:F0vsR_2temp}
\end{figure}

\noindent
As indicated by the results of the previous section, there is a need to find alternative procedures to reliably estimate the kinetic properties of the molecular motors when two different force-generating states are present. 
We here propose to extract such information by considering the effect of temperature on the properties of the motors. 
Before delving into the new fitting strategy, we outline how the temperature affects the properties of the system. 
It is commonly assumed, see e.g. \cite{piazzesi2003temperature}, that the magnitude $f_0$ of the force of a molecular motor in each of the attached state is not affected by the temperature changes. 
We here assume that its distribution does not change either, meaning that $f_1 \sim U(-f_0,f_0)$, $f_2 \sim U(\frac{f_0}{10},f_0)$ for all temperature of the system $T$.
Secondly, based on Arrenhius' theory of activated kinetic processes, the dependence of the reaction constants on the temperature of the system is assumed to be of the form:
\begin{equation}
	\label{kT}
	k_i(T_2)=k_i(T_1)Q_i^{\frac{\Delta T}{\SI{10}{\degreeCelsius}}}
\end{equation}
where $i = \{ \pm 1,\pm 2,3 \}$, $T_1$ and $T_2$ are two different temperatures, $\Delta T=T_2-T_1$ and the $Q_i$ represents the temperature coefficient for the $i-$th reaction. 
We remark that it is customary to measure the temperature differences in $\SI{10}{\degreeCelsius}$ and to refer to $Q_i$s as the  $Q_{10}$ factors. 
In the following we will indicate as $\bm{k}(T)$ the set of kinetic rates at temperature $T$. \\
We wish to investigate whether having data at two, or more, different temperatures helps with the estimation of the kinetic properties of the system. In order to do so, we consider two datasets $\{ F^{(1)}_i(t)\}_{i=1}^m$ and $\{ F^{(2)}_i(t)\}_{i=1}^m$ of synthetically generated trajectories of the total force at two different temperatures $T_1=\SI{25}{\degreeCelsius}$ and $T_2=\SI{15}{\degreeCelsius}$. 
Trajectories at temperature $T_1$ have been generated with a specified set of $\bm{k}(T_1)$ whereas the kinetic rates $\bm{k}(T_2)$ at temperature $T_2$ have been calculated through the relations \eqref{kT}, where we assumed the following values for the $Q_{10}$ factors: 
$Q_1= Q_{-1}= 2.0$, $Q_2= 5.5$, $Q_{-2}= 1.5$ and $Q_3=4.0$.
These values have been suitably chosen to mimic the temperature dependent performance of typical mammalian skeletal muscles as described in \cite{linari2007stiffness}.  
\\
In order to estimate the parameters $f_0$ and $\bm{k}(T_1)$ we propose the following modified optimisation procedure. A simulated annealing algorithm is applied to minimise the loss function
\begin{equation}
    \mathcal{L} = \mathcal{L}_1 + \mathcal{L}_2
\end{equation}
where the $\mathcal{L}_i$ are equivalent to the loss function described in the previous section and refer to the datasets at temperature $T_i$. We remark that not only $\mathcal{L}_1$ but also $\mathcal{L}_2$ is evaluated starting from the kinetic parameters $\bm{k}(T_1)$ by making use of equation \eqref{kT}.
The results of the fitting procedure performed on synthetic data at different temperatures are shown in Figure \ref{fig:fitDataT1T2} (a, for the force development) and (b, for the steady state force), while the comparison between the estimated parameters and the "true parameters', that is the ones set to generate the data, is provided in Table \ref{tab:parSim2temp} and Figure \ref{fig:parBestVSparTrue_2temp-lin}.
\begin{table}[h!]
	\begin{center}
		$\begin{array}{ccccccc}
			\toprule
			\text{} & F_0 \ (\SI{}{\pico \newton}) & f_0\ (\SI{}{\pico \newton}) & r & \phi (\SI{}{\second^{-1}})
			\\
			\midrule
			\text{True}& 16.9 & 3.0 & 0.68 & 1.500 &
			\\
			\text{parameters }&  &    &     & &\\
			& & & & & \\
			\text{Estimated }& 16.9 & 3.0 \pm 0.2 & 0.68 \pm 0.08 & 1.555 \pm 0.004  &
			\\
			\text{parameters }&   &   &     & &\\
			\bottomrule
		\end{array} $
	\end{center}
 \caption{\textbf{Estimated parameters via the inverse scheme fed with simulated data.}\\
		}
		\label{tab:parSim2temp}
\end{table}
\noindent
From the comparison between Figures \ref{fig:parBestVSparTrue} and  \ref{fig:parBestVSparTrue_2temp-lin} it is possible to see a remarkable improvement of the estimates of the kinetic constants, compared with the values obtained when only one single temperature dataset is available for analysis.
A better insight is gained by looking at the output of the optimisation procedure: see in Figure \ref{fig:F0vsR_2temp} the error bars of the symbol projected onto the parameter plane $(f_0,r)$ for either the one temperature (black square) or two (red diamond) temperatures. 
From equation \eqref{eq: First moment force} and the definition of the duty ratio we expect to see a shifted hyperbola given by
\begin{equation}
    r =  \frac{1}{f_0}\frac{20 F_0}{11N} + y^*. 
\end{equation}
Firstly, we observe that the assumption of high-temperature regime, adopted in \cite{buonfiglio2024force}, would result in $y*=0$ (blue dotted line in the Figure) and is clearly not adequate. The optimisation procedure proposed here is instead capable of handling virtually any temperature regime.
\\
Secondly, we remark that, if we only take into account the average profile of the ensemble force $F(t)$, we would not be able to resolve $f_0$ and $z(t)$ individually, as $F(t)$ depends on their product, see equation \eqref{eq: First moment force}. In this case, any point on the hyperbola would fit $F(t)$. On the other hand, accounting for the fluctuations of the force around the isometric plateau allows us to resolve this degeneracy and estimate $f_0$ and the duty ratio, as previously observed in \cite{buonfiglio2024force}.
\\
Thirdly, this visualisation shows that there is a clear improvement when using data coming from two different temperature settings (see red diamonds in Figure \ref{fig:F0vsR_2temp}) as compared to a single temperature dataset (black square). 
For reference, we show the true values set for generating the data as a yellow circle in Figure \ref{fig:F0vsR_2temp}. 
Having data at two different temperatures allows for reducing the degeneracy due to extreme large number of microscopic kinetic properties corresponding to same macroscopic ensemble force and, as a consequence, to better resolve rate dependent quantities such as average fraction of motors in $A_1$ configuration.

\section{Conclusions}
\noindent
The one-dimensional synthetic nanomachine described in \cite{pertici2018myosin} makes it possible to define the output of the interaction with an actin filament of a minimum number of myosin motors able to simulate the production of muscle contraction, namely generation of steady force and shortening by contractile proteins, in the absence of the confounding effects of regulatory and accessory proteins. 
In this way the nanomachine is able to reproduce the performance of reconstituted contractile systems powered by proteins purified from different muscle types. 
One main limit of the nanomachine is the presence of an intrinsically large trap compliance which causes any force generating interaction to reflect in sliding of the actin filament over the myosin array affecting the strain of the other attached motors. 
This effect has been mitigated by an implementation consisting in the possibility to switch the device to length clamp conditions \cite{buonfiglio2024force}.
In this way any interaction of the myosin array with the actin filament occurs in truly isometric conditions. 
Consequently, motors act as independent force generators and the force response is the result of the kinetics of attachment/force-generation/detachment events. 
For any interaction an initial growth of the force is recorded, followed by a subsequent saturation towards an asymptotic plateau. 
Around the asymptotic value, force fluctuations are displayed which contain sensible information on the inherent functioning of the examined molecular system.
In \cite{buonfiglio2024force}, a reverse engineering protocol was proposed and tested to infer a subset of key reference parameters from the experimental data. 
The proposed method was solely accounting for the contribution of attached motors in state $A_2$ (the so-called high force-generating state), which is the relevant attached state at the relatively high temperature of the experiment ($\SI{25}{\degreeCelsius}$). 
Under this condition, it was not possible to accurately resolve the kinetic reactions rates that characterised the cycle. 
In this paper, we take one decisive step forward by allowing for the contributions of both species $A_1$ and $A_2$ to be simultaneously accounted for, in the framework of an approximate solution of the stochastic model. 
This latter can be converted into a rather effective implementation of the fitting algorithm which performs extremely well against synthetically generated data. 
In this way, the newly proposed variant of the interpolation scheme can deal with multiple temperatures dataset to further constrain the phase space of searched parameters. 
By testing the fitting algorithm in silico, we demonstrate that the output of the nanomachine collected across a given temperature range can be fed to the model for simultaneous theoretical inspection providing the complete kinetic description of the collective motor under inspection. 
The challenging goal we plan to address is to apply the method against actual experimental data collected with the nanomachine powered not only by normal and mutant myosins, but also by engineered myosins, so as to  predict the corresponding muscle performance.
This integrated experimental and theoretical approach to investigate the molecular mechanisms governing the emergent collective mechanical output by an ensemble of muscle myosin motors, defines a topic of paramount importance in muscle biophysics and translational medicine.


\clearpage

\appendix

\clearpage
\section*{Appendix A: Thermodynamic consistency of the minimal kinetic model with an irreversible step} \label{App:TD}
\markboth{}{}

We consider an ensemble of $N$ myosin motors that performs isometric contractions. 
Each motor can be found in different configurations depending on its interaction with the actin filament: a detached state $D$ and two different force-generating attached states $A_1$ (low force configuration), and $A_2$ (high force configuration). 
Labelling as $n_D$ the number of detached motors, and as $n_i$, $i=1,2$ the number of attached motors in configurations $A_1$, and $A_2$ respectively, there is a natural conservation law for the number of motors in the system as:
\begin{equation}
	\label{consLaw}
	N= n_D+ n_1+n_2
\end{equation}
Each individual acto-myosin cyclic interaction is characterised by the following kinetic scheme:
\begin{center}
	\begin{tikzpicture}[x=0.75pt,y=0.75pt,yscale=-1,xscale=1]
		
		\draw[thin] (185,101.2) arc (20:90:16); 	
		\draw[thin] (237,155) arc (140:210:16); 	
		\draw[thin] (210,171) arc (220:290:16); 	
		
		\draw (133,107) node [anchor=north west][inner sep=0.75pt]    {${\displaystyle A_{1}}$};
		\draw (232,106.5) node [anchor=north west][inner sep=0.75pt]    {$A_{2}$};
		\draw (159,98.4) node [anchor=north west][inner sep=0.75pt]    {$\xrightleftharpoons[\quad k_{-2} \ \ ]{\ \ \quad \ k_{2} \quad}$};
		\draw (175,80) node [anchor=north west][inner sep=0.75pt]    {$Pi$};
		\draw (240,140) node [anchor=north west][inner sep=0.75pt]    {$ADP$};
		\draw (187,181.5) node [anchor=north west][inner sep=0.75pt]    {$D$};
		\draw (250,140) node [anchor=north west][inner sep=0.75pt]  [rotate=-127]  {$\xrightleftharpoons[ \quad k_{-3} \ \ \ ]{ \ \quad  k_{3} \quad}$};
		\draw (162,120) node [anchor=north west][inner sep=0.75pt]  [rotate=-53]  {$\xrightleftharpoons[\quad k_{1} \ \ ]{\quad k_{-1} \quad}$};
		\draw (230,170) node [anchor=north west][inner sep=0.75pt]  {$ATP$};
	\end{tikzpicture}
\end{center}
where $k_j$, $j=1,2,3$ denote the rate constants associated to each forward reaction channel, and $k_{-j}$ are the reverse rate constants.\\
Each step of this cyclic series of transitions between allowed configurations corresponds to a biochemical step that each motor undergoes during the ATPase cycle. \\
%
%
\begin{center}
\schemestart
\chemfig{\ce{A\cdot M\cdot ADP\cdot Pi <=> A\cdot M\cdot ADP +Pi}} \arrow(a--b){<=>[$k_{2}$][$k_{-2}$]} [,1.5,,thick] 
\chemfig{\ce{A\cdot M\cdot ADP <=> A\cdot M +ADP}}
\arrow(@a--c){<=>[${k_{-1}}$][${k_1}$]}[-45,1.5,,thick] 
\chemfig{ \ce{M\cdot ATP} } 
\chemfig{ \ce{<=>} } 
\chemfig{M\cdot ADP\cdot Pi} \hspace{-4cm}
\arrow(@b--e){<=>[$k_{-3}$][$k_3$]}[225,1.5,,thick] 
\schemestop
\end{center}
%
%
We define the following variables for the concentrations of the species of the model:
\begin{equation}
	\setlength{\jot}{5pt}
	\begin{array}{l l l l}
		\begin{cases}
			t &= [ATP] \\
			d &= [ADP] \\
			p &= [Pi]
		\end{cases}
		\qquad
		\begin{cases}
			x&= [D] \equiv [M \cdot ATP] (\equiv [M \cdot ADP \cdot Pi]) \\
			y&= [A_1] \equiv [A \cdot M \cdot ADP \cdot Pi]  \\
			z&= [A_2] \equiv [A \cdot M \cdot ADP] (\equiv [A \cdot M] )\\
		\end{cases}
	\end{array}
	\label{speciesConc}
\end{equation}
At the microscopic level the thermodynamic equilibrium corresponds to the principle of microscopic reversibility, that implied the detailed balance condition.
This condition is imposed by requiring that each individual reaction is balanced (i.e. each individual reaction occurs with equal forward and backward fluxes). 
A reaction system that satisfies detailed balance does not consume or dissipate free energy at thermodynamic equilibrium.
To impose this condition for the considered system implies a set of equations:
\begin{equation}
	\setlength{\jot}{5pt}
	\begin{cases}
		a \ x \ k_1 &= y \ k_{-1} \\
		y \ k_2 &= p \ z \ k_{-2} \\
		t \ z \ k_3 &= a \ x \ d \ k_{-3} \\  
	\end{cases}
\end{equation}
this implies that:
\begin{align*}
	\setlength{\jot}{8pt}
	& a \ x \ k_1 \ y \ k_2 \ t \ z \ k_r \ t \ z \ k_3 = y \ k_{-1} \ p \ z \ k_{-2} \ a \ x \ d \ k_{-3} \ ;\\
	& \frac{ k_1 \ k_2 \ k_3}{k_{-1} \ k_{-2} \ k_{-3} } = \frac{d \ p}{t} \\
\end{align*}
or: 
\begin{equation}
	\frac{ k_1 \ k_2 \ k_3}{k_{-1} \ k_{-2} \ k_{-3} } = \frac{[ADP]^{\text{eq}} \ [Pi]^{\text{eq}}}{[ATP]^{\text{eq}}} \ .
\end{equation}
From this condition it is possible to define of one of the rate constant in terms of the others as:
\begin{align*}
	\setlength{\jot}{10pt}
	k_{-3} \frac{[ADP]^{\text{eq}} \ [Pi]^{\text{eq}}}{[ATP]^{\text{eq}}} &= \frac{ k_1 \ k_2 \ k_3}{k_{-1} \ k_{-2} } 
\end{align*}
Since we are interested in experimental conditions that mimics isometric contractions in physiological [ATP] concentration we observe that the system is forced to perform in out of equilibrium conditions, specifically the [ATP] concentration is maintained constant at any time and the [ATP] hydrolysis products are continually extracted from the system, therefore:
\begin{equation}
	\frac{[ADP]^{\text{eq}} \ [Pi]^{\text{eq}}}{[ATP]^{\text{eq}}} > \frac{[ADP] \ [Pi]}{[ATP]} = \gamma
\end{equation}
We consider an effective non equilibrium rate constant $k_{-3}$ defined as:
\begin{equation}
	\hat{k}_{-3} \coloneqq 	\gamma k_{-3} < k_{-3}	\frac{[ADP]^{\text{eq}} \ [Pi]^{\text{eq}}}{[ATP]^{\text{eq}}} \ .
\end{equation}
For mammalian muscle fibre \cite{caremani2015force} we have:
\begin{equation}
	\gamma \simeq\frac{\SI{30}{\micro \molar} \times \SI{1}{\milli \molar}}{\SI{5}{\milli \molar}} \simeq 6 \times 10^{-6} \SI{}{\molar} \ .
\end{equation}
We can consider that the detachment step $\ce{A_2 <=> D}$ is effectively irreversible, and consequently set $k_{-3} \simeq 0$ when we study acto-myosin interactions in physiological ATP concentration. \\
To simplify the kinetic scheme it is possible to consider an effective description of the acto-myosin interaction cycle, by neglecting non equilibrium rate constant $\hat{k}_{-3}$ that combines the second order kinetic transition $k_{-3}$ and the factor $\gamma$, as well as the biochemical steps that solely involve the ATP and its hydrolysis products: 
\begin{equation}
\ce{ D <=>[\ce{\ k_1}\ ][\ce{\ k_{-1}}\ ] A_1 <=>[\ce{\ k_2 \ }][\ce{\ k_{-2}}\ ] \ce{A_2} ->[\ce{\ k_3 \ }][] D }
\end{equation}

\clearpage

\section*{Appendix B: Gaussian noise approximation for the motors populations}  \label{App:VK}
\markboth{}{}

We are now exploring another useful approximation of the dynamics of the stochastic system.
We are interested in an expression for the probability distribution of the populations dynamics, that involves the Gaussian noise approximation carried out on the master equation that describes the evolution of the probability distribution associated with the microscopic states of the system.
We observe that the marginal stationary solutions of the master equation $P^{\text{ST}}(n_1)$ and $P^{\text{ST}}(n_2)$, 
resemble normal distributions when the values of the fractions of motors are sufficiently far from the boundaries, i.e. for $n_1/N, n_2/N \gg 0$ and $n_1/N, n_2/N \ll 1$.
We then consider the probability distribution of the total force of the ensemble:
\begin{equation}
	\label{pdfTotalForce}
	P_{F_{n_1,n_2}}(\mathcal{F})= \sum_{q_1, q_2=0}^N P_{F_{q_1,q_2}} P(n_1=q_1, n_2=q_2)
\end{equation}
where $P_{F_{q_1,q_2}}(\mathcal{F}) \equiv \mathcal{G}\Bigl(\mu_F= \mu(q_1,f_0), \sigma_F^2=\sigma^2(q_1,q_2,f_0) \Bigr)$ is the Gaussian distribution obtained in Section \ref{sec: Distribution of force}.
In order to carry out the Gaussian noise approximation for the populations of motors we define the discrete number of motors in configuration $j$ as $n_j=Nx_j$ where $x_j$ is the "discrete" concentration of the motors in configuration $j$. 
Now we can express the stationary solution of the master equation in terms of the concentrations variables: $P(n_1=q_1, n_2=q_2)= P(Nx_1=q_1, Nx_2=q_2)$ or $P(x_1=\frac{q_1}{N}, x_2=\frac{q_2}{N})$.
In the limit $x_j \gg 1$ it is possible to perform a system size expansion of the master equation, a perturbative approach named the Van Kampen approximation \cite{van1976expansion}, in order to obtain a linear Fokker-Planck equation for the probability distribution of the fluctuations, around the mean field solution, associated with the concentration of the two force--generating motors populations. 
The stationary solution of the Fokker--Planck equation (i.e. the solution calculated when $\expval{\bm{x}}=\bm{x^*}$), written in terms of the concentration of motors is a bivariate Gaussian distribution, centred on the mean field solution of the dynamics of the system. 
The standard calculations carried out to obtain the Fokker--Planck equation and its solution from the master equation will be explicitly obtained in the next Section. 
This solution can be inserted in equation \eqref{pdfTotalForce} as an expression for $P(n_1=q_1, n_2=q_2)$ to be adopted instead of the exact solution of the master equation, which can be computationally expensive to be used in the fitting procedure.
The results of the parameter estimation on synthetic data sets will be presented after this Section.

\subsection*{Details of the Van Kampen expansion}

\begin{figure}[t]
	\begin{subfigure}{.5\textwidth}
		\includegraphics[scale=0.45]{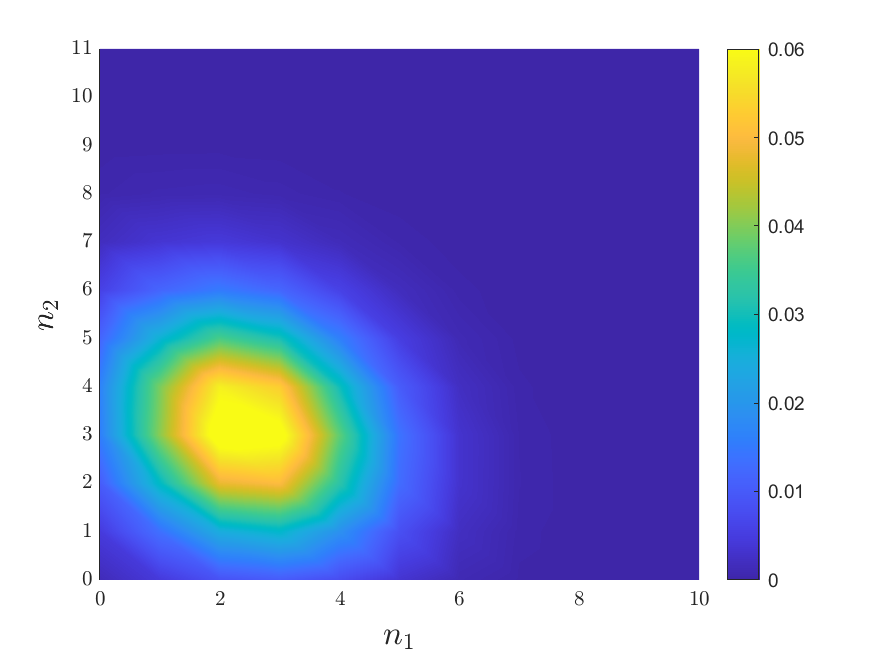}
		\subcaption{}
		\label{fig:a}
	\end{subfigure}
	\begin{subfigure}{.5\textwidth}
		\includegraphics[scale=0.45]{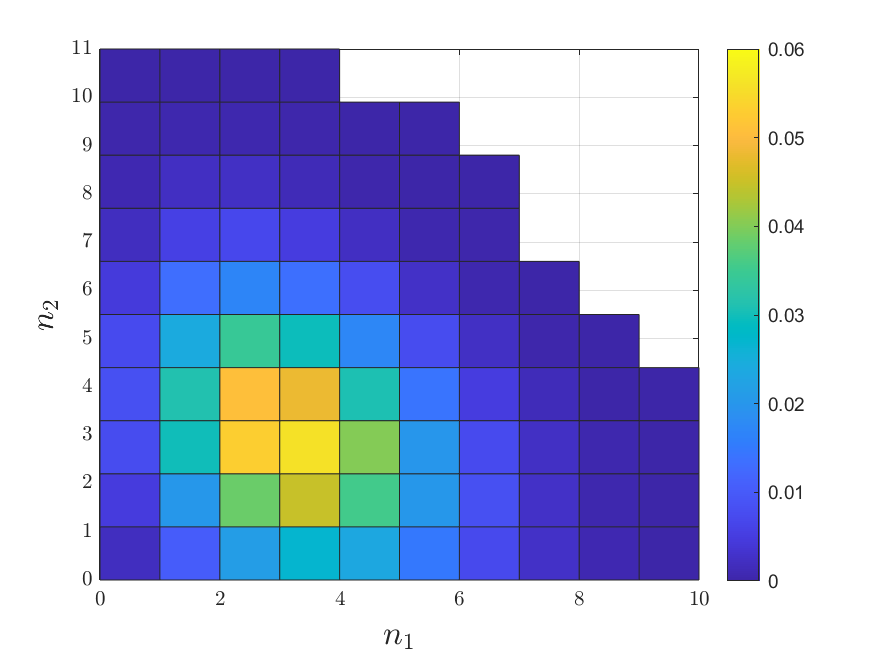}    
		\subcaption{}
		\label{fig:b}
	\end{subfigure}
	\caption{\textbf{Checking on the Van Kampen approximation.}\\
			Probability density function $P(n_1, n_2)$ associated to the stationary concentrations of force--generating motors, \textnormal{(a)} as obtained from stochastic simulations data, and \textnormal{(b)} from the analytical solution obtained via the Van Kampen approximation.
	}
	\label{fig:Pn1n2_gillVSvk}
\end{figure}

\noindent
To quantify the statistics of the fluctuations around the stationary state we consider the master equation in the form \eqref{eq: deterministic equations concentrations} and we perform an expansion in the system size $N$.
The first order of the expansion ($1/\sqrt{N}$) results in the means field equations for the populations of motors, while the second order of the expansion ($1/N$) yields a Fokker--Planck equation for the probability distribution of the finite size fluctuations around the stationary state, for each populations of motors.
Following the Van Kampen hypotheses, when the system size is large but finite $N\gg 1$, the discrete concentrations $n_1/N$ and $n_2/N$ will differ from the mean field fractions of motors by a contribution of magnitude $1/\sqrt{N}$ (as follows from the Central Limit Theorem):
\begin{equation}
	\label{VK}
	\frac{n_1(t)}{N}= y^*(t)+ \frac{\xi}{\sqrt{N}} \qquad \text{and} \qquad \frac{n_2(t)}{N}= z^*(t)+ \frac{\eta}{\sqrt{N}}
\end{equation}
where $y^*$ and $z^*$ are defined in equation \eqref{eq: mean field at any time}, and $\xi$ and $\eta$ are the fluctuations associated with the number of motors in $A_1$ and $A_2$ respectively.\\
%
To simplify the notation we recall that the number of motors in the actin--attached configurations at the time $t$ is indicated by $\bm{n}(t)=(n_1(t), n_2(t))$, and we define the vector containing the fractions of attached motors:
\begin{equation}
	\bm{x}= (y, z) \qquad \text{therefore} \qquad \bm{x^*}=(y^*,z^*)
\end{equation}
The associated fluctuations will be defined as: $\bm{\lambda}= (\xi, \eta)$. \\
These stochastic variables have a probability distribution $\Pi(\boldsymbol{\lambda}, t)$ defined by the following expression:
\begin{equation}
	\Pi(\boldsymbol{\lambda},t) \equiv P\bigl(\boldsymbol{n}; t\bigr)= P\Bigl(y+\frac{\xi}{\sqrt{N}}, z+\frac{\eta}{\sqrt{N}}; t\Bigr)
	\label{Pi}
\end{equation}
according to the Van Kampen hypothesis \eqref{VK}. \\ 
The time evolution of the probability distribution $\Pi(\boldsymbol{\lambda},t)$ is characterised by deriving the previous expression in respect of time:
\begin{equation}
	\setlength{\jot}{5pt}
	\begin{split}
		\frac{\partial P}{\partial t}&= \frac{1}{N} \frac{\partial \Pi}{\partial \tau}- \frac{1}{\sqrt{N}} \sum_{i=1}^2 {\frac{\partial\Pi}{\partial \lambda_i} \frac{dx_i}{d\tau}} = \\
		& = \frac{1}{\sqrt{N}} \Bigl( \frac{\partial P}{\partial t}\Bigr)_{1/\sqrt{N}}+ \frac{1}{N} \Bigl( \frac{\partial P}{\partial t}\Bigr)_{1/N}+ \mathcal{O} (N^{-3/2})
		\label{eq:master_Pi}
	\end{split}
\end{equation}
where we separated the two contributions of magnitude $1/\sqrt{N}$ and $1/N$. \\
To calculate these two contributions we write the master equations in terms of step operators  $\epsilon_i^\pm$ defined by their action:
\begin{equation}
	\epsilon_i^\pm T(n_i \vert n_i)P(n_i, t)= T(n_i\pm 1 \vert n_i\pm 1)P(n_i\pm 1, t)
\end{equation}
and, recalling the expressions for the transition rates we obtain:
\begin{equation}
	\label{ME_step}
	\setlength{\jot}{5pt}
	\begin{aligned}
		\frac{\partial P(\boldsymbol{n},t)}{\partial t} = &(\epsilon_1^{-} -1)T_1 P(\boldsymbol{n},t)+ (\epsilon_1^{+} -1)T_{-1} P(\boldsymbol{n},t)+ (\epsilon_1^{+} \epsilon_2^{-} -1)T_2 P(\boldsymbol{n},t) + \\
		& + (\epsilon_1^{-} \epsilon_2^{+} -1)T_{-2} P(\boldsymbol{n},t) + (\epsilon_2^{+} -1)T_3 P(\boldsymbol{n},t) \ .
	\end{aligned}
\end{equation}
We now expand both the step operators and the transition rates up to the second order ($\/N$), in the limit $1/\sqrt{N}\ll 1$. 
Observing that:
\begin{equation*}
	n_i \pm 1= \pm1 + N \Bigl( x_i+ \frac{\lambda_i}{\sqrt{N}} \Bigr)=  \pm 1+ N x_i+ \sqrt{N} \lambda_i 
\end{equation*}
we get:
\begin{equation*}
	\setlength{\jot}{5pt}
	\begin{aligned}
		\epsilon_i^\pm &\approx \mathbbm{1} \pm \frac{1}{\sqrt{N}} \partial_{\lambda_i}+ \frac{1}{2N} \partial_{\lambda_i}^2+ \mathcal{O}(N^{-3/2}) \\
		& \\
		\epsilon_i^{\pm} \epsilon_j^{\mp} &\approx \mathbbm{1} \mp \frac{1}{\sqrt{N}} \partial_{\lambda_j}+ \frac{1}{2N} \partial_{\lambda_j}^2  \pm \frac{1}{\sqrt{N}} \partial_{\lambda_i}- \frac{1}{N} \partial_{\lambda_i}\partial_{\lambda_j} + \frac{1}{2N} \partial_{\lambda_i}^2+ \mathcal{O}(N^{-3/2}) \approx \\
		& \approx \mathbbm{1}+ \frac{1}{\sqrt{N}}(\pm \partial_{\lambda_i} \mp \partial_{\lambda_j})+ \frac{1}{2N}(\partial_{\lambda_i}^2- \partial_{\lambda_j}^2+ 2\partial_{\lambda_i\lambda_j}^2) \\
	\end{aligned}
\end{equation*}
and for the transition rates:
\begin{equation*}
	\setlength{\jot}{5pt}
	\begin{aligned}
		T_1  \approx  k_1 \Bigl [ 1- y- z - \frac{(\xi+ \eta)}{\sqrt{N}} \Bigr ] \qquad &\text{and} \qquad 	T_{-1}  \approx  k_{-1} \Bigl(y+ \frac{\xi}{\sqrt{N}} \Bigr) \\ 
		T_2  \approx  k_2 \Bigl(y+ \frac{\xi}{\sqrt{N}} \Bigr) \qquad &\text{and} \qquad 	T_{-2}  \approx  k_{-2} \Bigl(z+ \frac{\eta}{\sqrt{N}} \Bigr) \\ 
		T_3  \approx  k_3 \Bigl(z+ \frac{\eta}{\sqrt{N}} \Bigr)  \qquad &
	\end{aligned}
\end{equation*}
Substituting these expressions in the master equation \eqref{ME_step} and comparing the results with the equation \eqref{eq:master_Pi}, we find that the first term on the right hand side (the leading order of the expansion) coincides with the set of differential equations that governs the mean field dynamics of the system.
The second term on the right hand side results to be: 
\begin{equation*}
	\label{eq:second_ord}
	\setlength{\jot}{5pt}
	\begin{aligned}
		\Bigl( \frac{\partial P}{\partial t}\Bigr)_{1/N} = \frac{\partial \Pi}{\partial t}=& - \partial_{\xi}(A_1 \Pi(\lambda;t))-\partial_{\eta}(A_2  \Pi(\lambda;t))+
		\frac{1}{2} \Bigl [ \partial^2_{\xi ^2} (B_{11} \Pi(\lambda;t))+ \\
		& + \partial^2_{\eta ^2} (B_{22} \Pi(\lambda;t))+  
		\partial_{\xi}  \partial_{\eta} (B_{12}  \Pi(\lambda;t)) + \partial_{\eta}  \partial_{\xi} (B_{21}  \Pi(\lambda;t))\Bigr ]  
	\end{aligned}
\end{equation*}
where $\bm{A}(\boldsymbol{\lambda})=M \bm{\lambda}^T$. 
Matrices M and B, respectively the drift matrix and the diffusion matrix are:
\begin{equation}
	M=
	\begin{pmatrix}
		-(k_1+k_{-1}+k_2) &\quad -(k_1-k_{-2}) \\
		& & \\
		k_2 &\quad -(k_{-2}+k_3)
	\end{pmatrix}
	\label{Mmatrix}
\end{equation}
which coincides in the deterministic limit, with the jacobian matrix of the system \eqref{Jmatrix}, and: 
\begin{equation}
	B=
	\begin{pmatrix}
		k_1(1-y-z)+ (k_{-1}+k_2) y+ k_{-2}z &\quad -k_2 y+ k_{-2} z \\
		& & \\
		-k_2 y+ k_{-2} z &\quad k_2 y + (-k_2+ k_3) z
	\end{pmatrix}
	\label{Bmatrix}
\end{equation}
which is a symmetric and positive definite matrix. \\
The previous expression can be written as standard Fokker--Plank equation in the form:
\begin{equation}
	\label{FP}
	\frac{\partial \Pi(\boldsymbol{\lambda},t)}{\partial t}= - \sum_{i=1} {\frac{\partial}{\partial \lambda_i} \Bigl[A_i(\boldsymbol{\lambda}) \Pi(\boldsymbol{\lambda}, t)\Bigr]}+ \frac{1}{2} \sum_{i,j=1}^2 {\frac{\partial^2}{\partial \lambda_i \partial \lambda_j} \Bigl[B_{ij}(\boldsymbol{\lambda}) \Pi(\boldsymbol{\lambda}, t )\Bigr]} \ .
\end{equation}
The Van Kampen approximation allowed us to decouple the deterministic and the stochastic dynamics of the system, which results in local fluctuations around the mean field stationary state. 
Summarising, the second order in $\sqrt{N}$ of the expansion yields a Fokker--Planck equation for the probability distribution of the fluctuations, associated with the populations of motors in the actin--attached configurations.
The general solution of the Fokker-Planck equation \eqref{FP} \cite{van1976expansion}, is a bivariate Gaussian distribution, that can be characterised by its moments.
Performing the standard calculations, the equation that describes the time evolution of he first moment of the distribution is, for each component, the linear differential equation is: 
\begin{equation}
	\frac{d \expval{\lambda_i}}{d \tau}= \sum_{k=1}^2{M_{ik} \expval{\lambda_k}}= \expval{A_i}
\end{equation}
or, in components:
\begin{equation}
	\label{firstMoment}
	\setlength{\jot}{5pt}
	\begin{split}
		\dot{\expval{\xi}}&= \expval{A_1}=  \sum_{k=1}^2{M_{1k}\expval{\lambda_k}}=  (k_1+k_{-1}+k_2)\expval{\xi}- (k_1-k_{-2})\expval{\eta} \\
		\dot{\expval{\eta}}&= \expval{A_2}= \sum_{k=1}^2{M_{2k}\expval{\lambda_k}}= k_2 \expval{\xi}- (k_{-2}+k_3) \expval{\eta} \\
	\end{split}
\end{equation}
that implies that for $\dot{\expval{\bm{\lambda}}}=0$ we have $\expval{\lambda_i}=0$ for $i=1,2$, accordingly with the Van Kampen ansatz.\\
For the second moment of the distribution we obtain the following set of differential equations, for the diagonal components:
\begin{equation}
	\frac{d \expval{\lambda_i^2}}{d\tau}= 2 \sum_{k=1}^2{M_{ik} \expval{\lambda_i\lambda_k}}+ B_{ii}= 2 \expval{\lambda_i A_i}+ B_{ii} 
\end{equation}
and for the off-diagonal ones:
\begin{equation}
	\setlength{\jot}{5pt}
	\begin{aligned}
		\frac{d \expval{\lambda_i\lambda_j}}{d\tau}&= \sum_{k=1}^2{\Bigl(M_{ik} \expval{\lambda_j\lambda_k}+ M_{jk} \expval{\lambda_j\lambda_k}\Bigr)}+ \frac{1}{2}B_{ij}+ \frac{1}{2}B_{ji}= \\
		& = \expval{\xi_j A_i}+ \expval{\xi_i A_j}+ \frac{1}{2} (B_{ij}+ B_{ji} ) 
	\end{aligned}
\end{equation}
or, in components:
\begin{equation}
	\setlength{\jot}{5pt}
	\begin{aligned}
		\dot{\expval{\xi^2}} = 2 \sum_{k=1}^2 {M_{1k} \expval{\xi \lambda_k}}&+ B_{11} \qquad \qquad  		\dot{\expval{\eta^2}}= 2 \sum_{k=1}^2 {M_{2k} \expval{\eta \lambda_k}}+ B_{22} \\
		\dot{\expval{\xi \eta}}&= \sum_{k=1}^2 {M_{1k} \expval{\eta \lambda_k}+ M_{2k} \expval{\xi \lambda_k}}+ B_{12} \\
	\end{aligned}
\end{equation}
where $B_{12}=B_{21}$.\\
We now define the vector of the second moments of the distribution as $\bm{\zeta}= \Bigl ( \expval{\xi^2} \quad \expval{\eta^2} \quad \expval{\xi \eta} \Bigr )^T$, that satisfies the following equation for the time evolution:
\begin{equation}
	\dot{\expval{\bm{\zeta}}}= \mathcal{M} \expval{\bm{\zeta}}+ \bm{b}
\end{equation}
where $\bm{b}= (B_{11} \quad B_{22} \quad B_{12})^T$, and $\mathcal{M}$ is the matrix with the following elements:
\begin{equation}
	\mathcal{M}=
	\begin{pmatrix}
		2 J_{11} &\quad 0 &\quad 2 J_{12}\\
		0 &\quad 2 J_{22} &\quad 2 J_{21} \\
		J_{21} &\quad J_{12} &\quad (J_{11}+ J_{22})
	\end{pmatrix}
	\label{zetaMatrix}
\end{equation}
Being interested in the solution of the previous equation for the dynamics of the fluctuations around the stationary state we are interesting to solve the previous equation when $\dot{\expval{\bm{\zeta}}}=0$, i.e.:
\begin{equation}
	\expval{\bm{\zeta}}=-\mathcal{M}^{-1}\bm{b}	\ .
\end{equation}
The solution of the Fokker--Planck equation \eqref{FP} has the explicit form:
\begin{equation}
	\Pi(\bm{\lambda};t)=  \frac{\exp(- \frac{1}{2} \Bigl [ (\bm{\lambda}- \expval{\bm{\lambda}})^T \Sigma ^{-1} (\bm{\lambda}- \expval{\bm{\lambda}})\Bigr ])}{2 \pi \sqrt{det(\Sigma)}}
\end{equation}
where $\expval{\bm{\lambda}}= (\expval{\xi} \quad \expval{\eta})^T$ is the mean value of the vector of the fluctuations of the two motors populations, and $\Sigma$ is the covariance matrix with elements:
\begin{equation}
	\Sigma=
	\begin{pmatrix}
		\expval{\xi^2} &\quad \expval{\xi \eta}  \\
		\expval{\xi \eta} &\quad \expval{\eta^2}
	\end{pmatrix}
	\label{SigmaMatrix}
\end{equation}
Equation \eqref{FP} can be expressed in terms of the components of the fluctuations associated with the two motors population as:
\begin{equation}
	\Pi(\xi, \eta; t)=  \frac{\exp(- \frac{1}{2(1- \rho^2)} \Bigl [ \frac{( \xi- \expval{\xi} )^2}{\expval{\xi^2}}- 2 \rho \frac{(\xi- \expval{\xi}) (\eta- \expval{\eta})}{\sqrt{\expval{\xi^2} \expval{\eta^2}}} + \frac{(\eta- \expval{\eta})}{\expval{\eta^2}} \Bigr ] )}{2 \pi \sqrt{\expval{\xi^2} \expval{\eta^2}(1-\rho^2)}}
\end{equation}
where the correlation coefficient $\rho$ is defined as:
\begin{equation}
	\rho= \frac{\expval{\xi\eta}}{\sqrt{\expval{\xi^2}\expval{\eta^2}}} \ .
\end{equation}
We are interested in the solution of the Fokker--Planck equation expressed for the number of motors in the actin--attached configurations, i.e. $n_1$ and $n_2$, which is:
\begin{equation}
	\label{Pn1n2_VK}
	P(n_1, n_2)= \frac{\exp ( - \frac{1}{2 (1-\rho ^2) N} \Bigl [ \frac{(n_1- N y^*)^2}{\expval{\xi^2}}- 2\rho \frac{(n_1- N y^*)(n_2- N z^*)}{\sqrt{\expval{\xi^2}\expval{ \eta^2}}} + \frac{(n_2- N z^*)^2}{\expval{\eta^2}}\Bigr ]) }{2 \pi \sqrt{N^2 \expval{\xi^2} \expval{\eta^2} (1-\rho^2)}}
\end{equation}
where we have exploited that $\expval{\xi}= \expval{\lambda}=0$ from \eqref{firstMoment}, $\xi= \frac{(n_1-N y^*)}{\sqrt{N}}$ from the Van Kampen hypotheses \eqref{VK} and that $\Sigma_{\bm{\lambda}}= \frac{\Sigma_{\bm{n}}}{N}$. \\
Before we proceed to utilise this expression in the definition of the probability distribution of the force, we must inspect if this approximation is satisfying for the conditions under which we performed the stochastic simulations of the system dynamics. 
In Figure \ref{fig:Pn1n2_gillVSvk} we can see the $2D$ histogram of the stationary probability distribution $P(n_1,n_2)$ obtained from the Gillespie simulation of a stochastic trajectory (on the left), and the corresponding function derived by the implementation of the Van Kampen approximation on the master equation (on the right).
The agreement between the two results is satisfactory when we consider a set of kinetic rates that correspond to average fractions of motors that are sufficiently different from zero in the stationary state, i.e. $\expval{n_1},\expval{n_2} \gg 0$.  \\
\noindent
By inserting expression \eqref{eq: PF_n1n2-VK} in the probability distribution of the total force of the ensemble \eqref{eq: PF_n1n2} (replacing the exact solution of the master equation), when considering the continuous limit for the concentrations, we find:
\begin{equation}
	\label{eq: PF_n1n2-VK}
	\setlength{\jot}{5pt}
	\begin{aligned}
		P_{F_{n_1,n_2}}(\mathcal{F}) &= \int_0^N dq_1 \int_0^N dq_2 \, P_{F_{q_1,q_2}} (\mathcal{F}) \, P(Ny=q_1, Nz=q_2)=\\
		=& \int_0^N dq_1 \int_0^N dq_2 \, \mathcal{G}\Bigl(\bm{\mu}_F, \bm{\sigma_F}\Bigr) \, \mathcal{G}\Bigl( \bm{x}^*, \frac{1}{N}\bm{\Sigma} \Bigr)= \\
		&= \int_{- \infty}^{+ \infty} dx_1 \int_{- \infty}^{+ \infty} dx_2 \, \frac{e^{-\frac{(\mathcal{F}- \mu_F)^2}{2\sigma_F^2}}}{\sqrt{2 \pi \sigma_F^2}} \, \frac{e^{-\frac{1}{2N}(\bm{x}- \bm{x^*})^T \Sigma^{-1} (\bm{x}- \bm{x^*})}}{2 \pi \sqrt{N^2 \text{det}(\Sigma)}} 
	\end{aligned}
\end{equation}
The integral can be numerically evaluated and exploited to perform a fitting procedure that does not involve the calculations needed to solve the master equation. \\
The results of this procedure will be presented in the next Section.

\clearpage
\subsection*{Fitting scheme for data at different temperatures}

\begin{figure}[!t]
	\centering
	\includegraphics[scale=0.8]{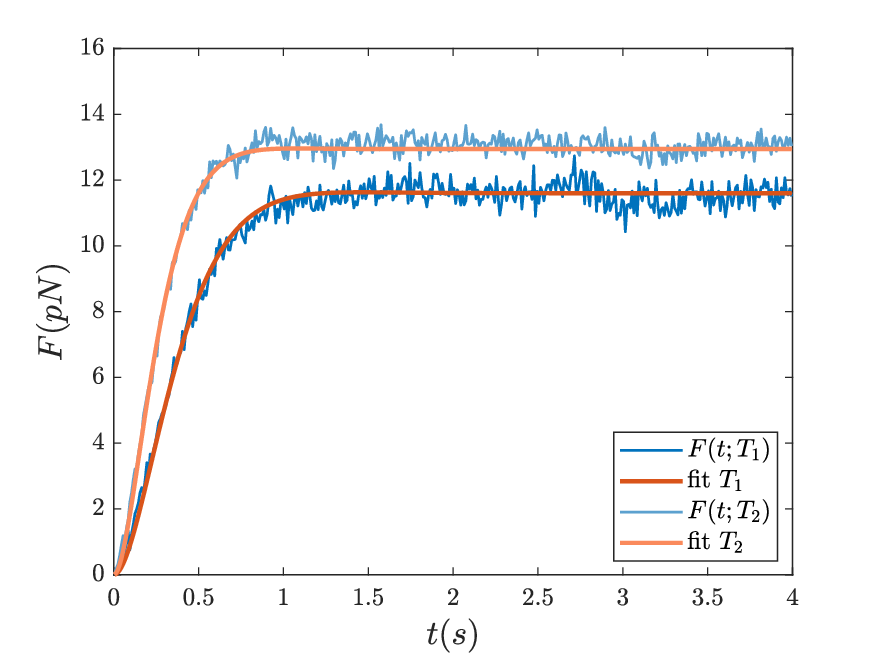}
	\caption{\textbf{Result of the fitting procedure of the force development performed at two different temperatures.}\\
			The trajectory of the force has been obtained averaging over $800$ independent simulated trajectories. 
			The force is measured in pN and it is exerted by a collection of $N = 20$ molecular motors with a suitable choice of the kinetic parameters $\bm{k}$ and $f_0$.
			The relevant parameters of the simulations are: $N=20$, $f_0=3, T_1= \SI{10}{\degreeCelsius}$ (darker colours) and $T_2= \SI{14}{\degreeCelsius}$ (lighter colours).
	}
	\label{fig:fitDataDevT1T2-VK}
\end{figure}

\begin{figure}[t!]
	\centering
	\includegraphics[scale=0.8]{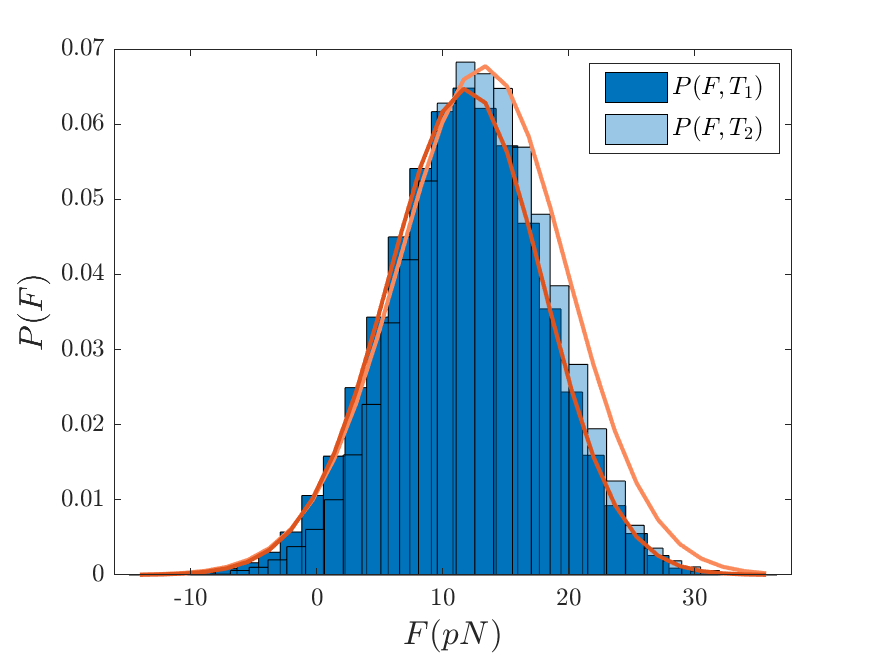} 
	\caption{\textbf{Result of the fitting procedure on the probability distribution of the force obtained from the stochastic simulations performed at two different temperatures.}\\
			The histogram of the force of the ensemble at the isometric plateau is fitted against the analytical profile, via a self–consistent optimisation procedure which aims at estimating the kinetic parameters and the force of a single motor $f_0$. The relevant parameters of the simulations are: $N=20$, $f_0=3, T_1= \SI{10}{\degreeCelsius}$ (darker colours) and $T_2= \SI{14}{\degreeCelsius}$ (lighter colours). 
	}
	\label{fig:fitDataT1T2-VK}
\end{figure}

\begin{figure}[!t]
	\centering
	\includegraphics[scale=0.8]{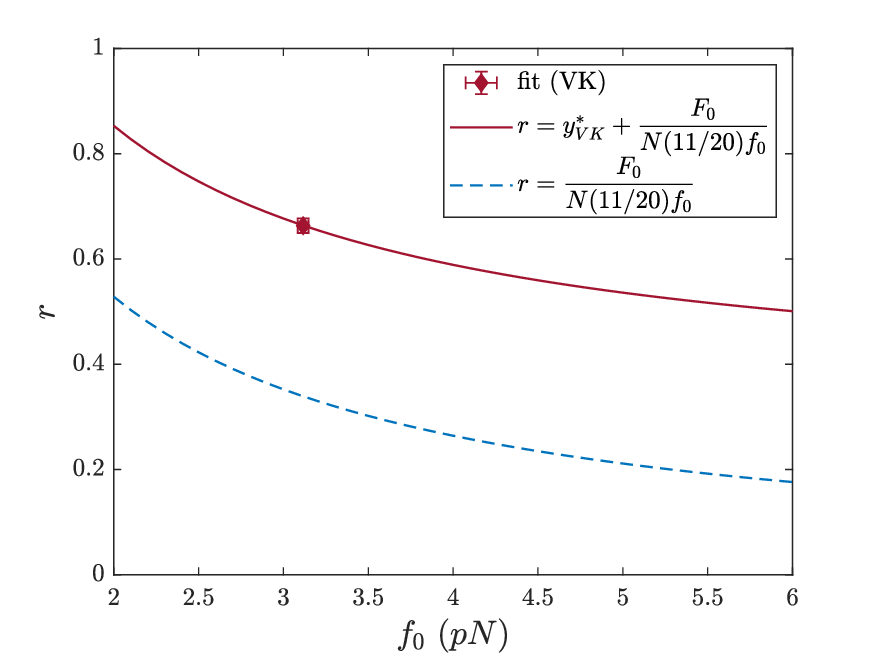} 
	\caption{\textbf{Estimated parameters $(\bm{f_0}, \bm{r})$ from the synthetic data.}\\
			The red symbol (mean $\pm$ SD) represents the solution of the optimisations procedure for the force of a single motors $f_0$, and the duty ratio of the ensemble $r$. 
			The solid red line represents the expression of the duty ratio of the ensemble as it follows from the mean field model, with the mean field solution for $y^*$ expressed as a function of the kinetic parameters of the system. 
			The dashed blue line represents the duty ratio as previously estimated without taking into account the contribution of the motors in configuration $A_1$. 
	}
	\label{fig:F0vsR_2temp-VK}
\end{figure}

\noindent
In this Section we show the results of the new approach that exploits the functional form \eqref{eq: PF_n1n2-VK}  for the probability distribution of the total force of the ensemble to perform the parameter estimation on data sets numerically generated with different temperature parameters for which the Van Kampen approximation can be carried out. 
In Figure \ref{fig:fitDataDevT1T2-VK} and in Figure \ref{fig:fitDataT1T2-VK} are shown the results of the fitting procedure on the force development and on the probability distribution of the force of the ensemble.
The average values of the parameters obtained with this method are in good accord with the parameters adopted to generate the simulated trajectories, as can be appreciated inspecting the results reported in Table \ref{tab:parSim2temp_VK}. \\
\begin{table}[h!]
	\begin{center}
		\caption{\textbf{Estimated parameters via the inverse scheme fed with simulated data.}\\
		}
		\label{tab:parSim2temp_VK}
		$\begin{array}{ccccccc}
			\toprule
			\text{} & F_0 \ (\SI{}{\pico \newton}) & f_0\ (\SI{}{\pico \newton}) & r & 
			\\
			\midrule
			\text{True}& 11.6 & 3.0 & 0.70 & 
			\\
			\text{parameters }&  &    &     & &\\
			& & & & & \\
			\text{Estimated }& 11.6 & 3.12 \pm 0.2 & 0.66 \pm 0.02 & 
			\\
			\text{parameters }&   &   &     & &\\
			\bottomrule
		\end{array} $
	\end{center}
\end{table}
\\
\noindent
In Figure \ref{fig:F0vsR_2temp-VK} it is shown the result of the fitting procedure in the plane $(f_0,r)$, where the solutions of the mean field dynamics are represented by the two hyperbolae, the one above corresponding to the data set with the lower temperature $T_1$, while the dashed blue line represents the duty ratio as previously estimated without taking into account the contribution of the motors in configuration $A_1$. 
The symbols are the results of the analysis of the probability distribution of the force fluctuations around the isometric plateau, exerted by the ensemble of motors.

\clearpage
\printbibliography[heading=bibintoc]

\clearpage
\markboth{}{}

\noindent
\textbf{Acknowledgments}\\
This work was supported by the European Joint Program on Rare Diseases 2019 (PredACTINg, EJPRD19–033). NZ has been supported by the Wallenberg Initiative on Networks and Quantum Information (WINQ)
\\

\end{document}